\documentclass[10pt,a4paper]{article}
\usepackage[utf8]{inputenc}
\usepackage[T1]{fontenc}
\usepackage{booktabs}
\usepackage{amssymb}
\usepackage{graphicx}
\usepackage{xcolor}
\usepackage{mathtools}
\usepackage{amsthm}
\usepackage{comment}
\usepackage{amsbsy}
\usepackage{subcaption}
\usepackage{geometry}
\usepackage{color}
\usepackage{physics}
\usepackage{soul}
\usepackage{cancel}
\usepackage{float}
\usepackage{jheppub}
\usepackage{cleveref}

\newcommand{\GeV}{\text{GeV}}
\newcommand{\mAp}{m_{A^\prime}}
\newcommand{\mcc}{m_{\chi \chi}}
\newcommand{\Eb}{\frac{s-{\mAp^2}}{2 \sqrt{s}}}

\title{Dark Higgs-strahlung at Belle~II: A distinctive dark sector signature with displaced vertices and missing energy }
\author[1,2]{Francesca Acanfora}
\author[1,3]{and Felix Kahlhoefer}
\affiliation[1]{Institute for Theoretical Particle Physics (TTP), Karlsruhe Institute of Technology (KIT), \\D-76131 Karlsruhe, Germany}
\affiliation[2]{Department of Physics, University of Massachusetts, Amherst, MA 01003, USA}
\affiliation[3]{Institute for Astroparticle Physics (IAP), Karlsruhe Institute of Technology (KIT), Hermann-von-\\Helmholtz-Platz 1, 76344 Eggenstein-Leopoldshafen, Germany}

\emailAdd{facanfora@umass.edu}
\emailAdd{kahlhoefer@kit.edu}

\abstract{
Dark photons with kinetic mixing are compelling mediators for the interactions between dark matter and Standard Model particles. While most experimental searches focus on fully visible or fully invisible decays of dark photons, we explore processes that involve dark Higgs-strahlung, i.e.\ the emission of a dark Higgs boson connected to the mass generation of the dark photon. If the dark Higgs boson is the lightest dark sector particle, it is expected to be long-lived and decay into Standard Model particles via Higgs mixing. At electron-positron colliders, dark Higgs-strahlung may occur either in isolation (leading to a single displaced vertex and missing energy) or accompanied by a photon from initial-state radiation. Both signatures offer distinctive kinematic features, such as peaks in photon energy or missing invariant mass, which enable efficient background suppression and enhances sensitivity beyond existing searches. Our study suggests that Belle~II could significantly improve coverage of dark sector models by targeting this previously unexplored final state and that combining dark Higgs-strahlung events with and without additional photon offers great potential for reconstructing the properties of the dark sector.}


\preprint{P3H-25-056, TTP25-025}

\graphicspath{{plots}}

\begin{document}

\maketitle
\flushbottom

\section{Introduction}

Dark photons arising as massive vector bosons from a spontaneously broken $U(1)'$ gauge symmetry are one of the simplest and most compelling examples for feebly-interacting particles, i.e.\ particles with small mass compared to the electroweak scale and tiny couplings to Standard Model (SM) states~\cite{Antel:2023hkf}. For the simplest dark photon model, in which SM particles do not carry $U(1)'$ charge, interactions with known particles arise only via kinetic mixing with the hypercharge gauge boson~\cite{Holdom:1985ag}, leading to a highly predictive scenario that has been the target of a wide range of experimental searches~\cite{Fabbrichesi:2020wbt}. If the dark photon decays visibly, it can be targeted for example with searches for di-muon resonances, arising either from a prompt or displaced vertex, see e.g. Ref.~\cite{LHCb:2020ysn}. If, on the other hand, the dark photon decays into a pair of invisible particles (such as dark matter), it can be probed with searches for missing energy, for example at electron-positron colliders such as BaBar or Belle~II. The simplest example for such a search is a single-photon search, targeting processes in which a photon from initial-state radiation  (ISR) recoils against an invisibly decaying dark photon. The known kinematics of the process lead to a peak in the photon energy distribution that can be used to reconstruct the dark photon mass without observing its decay products~\cite{BaBar:2017tiz}.

\begin{figure}[H]
\centering
\includegraphics[width=0.48\textwidth,clip,trim=0 -10 0 0]{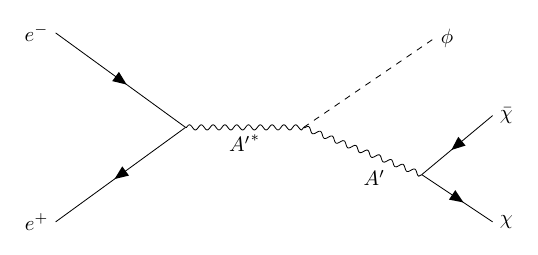}
\hfill
\includegraphics[width=0.48\textwidth,clip,trim=0 10 0 0]{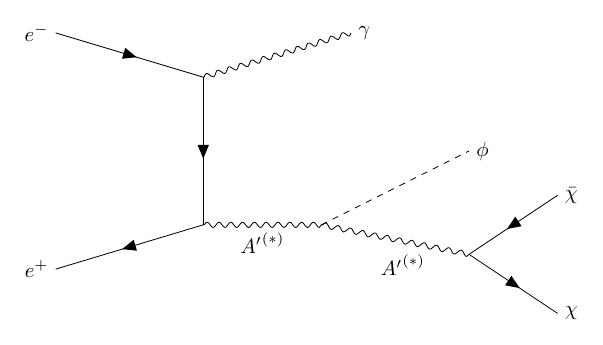}
\caption{Feynman diagrams for the two dark Higgs boson production processes considered in this work. We refer to the left process as $\phi$ production and to the right as $\gamma \phi$ production.}
\label{fig:fey_diags}
\end{figure}

While often left unspecified, an important ingredient for realistic dark photon models is the mass generation mechanism. The most well-motivated possibility is to generate the dark photon mass through a dark Higgs mechanism in analogy with electroweak symmetry breaking~\cite{Schabinger:2005ei}. In this case we expect the dark Higgs boson to be comparable in mass to the dark photon, such that it can be produced in the same processes that also produce dark photons. In particular, dark Higgs bosons can be radiated from the dark photon (so-called dark Higgs-strahlung~\cite{Duerr:2017uap}), in analogy to $Zh$ production at high-energy electron-positron colliders, see the left panel of figure~\ref{fig:fey_diags}. If the spectrum of the dark sector is such that the dark photon decays invisibly but the dark Higgs boson decays into SM particles (via mixing with the SM Higgs boson), the resulting signature would be a pair of muons or hadrons with an invariant mass peaking at the dark Higgs boson mass, accompanied by an missing energy peak similar to that in single-photon searches.

In spite of its conceptual simplicity, this final state has so far not been targeted by electron-positron colliders. This is in contrast to the LHC, where the dark Higgs-strahlung process has been studied in the $b\bar{b}$~\cite{ATLAS:2024ypx}, $W^+W^-$~\cite{ATLAS:2022bzt,CMS:2023dof}, $ZZ$~\cite{ATLAS:2020fgc} and $hh$~\cite{ATLAS:2025auy} final states, even though in this case the missing energy distribution is much broader and does not allow for a direct reconstruction of the dark photon mass. The reason for this gap in our coverage of dark sectors is that for dark photon masses at the GeV scale the rather unspecific final state makes both triggering and background rejection challenging.\footnote{Note, however, that Belle~II has recently carried out a search for dark Higgs bosons produced in association with a semi-visibly decaying dark photon, a signature that arises naturally in models of inelastic dark matter~\cite{Belle-II:2025bhd}. Moreover, there exist various searches for light scalars with Higgs mixing that do not rely on the presence of a dark photon, for example in rare $B$ meson decays~\cite{Belle-II:2023ueh}.}

In the present work we point out a promising alternative: dark Higgs-strahlung combined with an ISR photon, see the right panel of figure~\ref{fig:fey_diags}. This process has recently been considered for the case of an invisibly decaying dark Higgs boson in Ref.~\cite{Li:2025tlg}, whereas we consider the case of visible decays. The cross section for this process is enhanced compared to the naive expectation, because the dark photon can be on-shell or nearly on-shell before emitting the dark Higgs boson. However, the kinematics of the process is now ambiguous: if the photon is on-shell before emitting the dark Higgs boson, we expect a peak in the energy of the single photon and a broad distribution of missing invariant mass, whereas a broad distribution of single-photon energies and a peak in the missing invariant mass is expected if the dark photon is on-shell after emitting the dark Higgs boson. 

We show that these different features can be combined into a single kinematic variable that can be used to efficiently suppress background. Combined with the displaced vertex from the visible dark Higgs boson decay, this yields a striking signature that offers a promising target for Belle~II. Using a simple mock background distribution, we estimate the sensitivity that can be achieved and show that dark Higgs boson searches may significantly surpass single photon searches even for large background rates. Moreover, detecting dark Higgs bosons both with and without an ISR photon would give additional information about the underlying process that can be used to accurately reconstruct the dark sector properties. 

We emphasize that the model set-up we consider in this work differs slightly from previous studies, because we assume that the dark matter particle does not couple to the dark Higgs boson and obtains its mass independently. The key advantage of this set-up, apart from the simpler kinematics, is that it provides a new channel for setting the dark matter relic abundance: the annihilation into dark Higgs bosons via dark photon loops. We calculate these loops and find that the resulting annihilation process can reproduce the dark matter relic abundance for gauge couplings of order unity across the entire parameter space that we consider. This result adds further motivation to the exploration of dark Higgs-strahlung.

The remainder of this work is structured as follows. In section~\ref{sec:model} we introduce the model that we study and the benchmark scenarios that we will use for our analysis. The simulation of signal and mock background are described in section~\ref{sec:simulation}. The distributions and relevant kinematic variables are discussed in section~\ref{sec:distributions}. Finally, we present and discuss our results in section~\ref{sec:results} before concluding in section~\ref{sec:conclusions}.

\section{Model set-up}\label{sec:model}

We consider a dark sector comprising three particles: a dark photon $A'$, a dark Higgs field $\Phi$ and a Dirac fermion dark matter particle $\chi$. The dark Higgs field carries charge $q_\Phi = 1$ under the $U(1)'$ gauge symmetry with gauge coupling $g^\prime$, such that its interactions are given by
\begin{equation}
\mathcal{L}_\Phi = \left[ \left( \partial^\mu + i g^\prime q_\Phi A^{\prime \mu} \right) \Phi\right]^\dagger\left[ \left( \partial_\mu + i g^\prime q_\Phi A^\prime_\mu \right) \Phi\right] - V(\Phi, H) \; .
\end{equation}
The scalar potential $V(\Phi, H)$ induces spontaneous symmetry breaking, giving the dark Higgs field a vacuum expectation value $w$:
\begin{equation}
\Phi = \frac{\phi + w}{\sqrt{2}}
\end{equation}
with the physical dark Higgs boson $\phi$. The potential also includes the portal interaction $\lambda_{h\phi} |\Phi|^2 |H|^2$, leading to mixing between the two Higgs bosons, such that
\begin{align}
h & \to \cos \theta \, h + \sin \theta \, \phi \\
\phi & \to - \sin \theta \, h + \cos \theta \, \phi \; ,
\end{align}
with the mixing angle $\theta \ll 1$.

The breaking of the $U(1)'$ symmetry gives the dark photon a mass $m_{A'} = q_\Phi g^\prime w$. For the dark matter particle $\chi$, we assume vector-like couplings to the dark photon (i.e.\ $q_{\chi_L} = q_{\chi_R}$), such that a Yukawa interaction with the dark Higgs boson is forbidden. Instead, the mass of the dark matter particle is given directly by a Dirac mass term:
\begin{equation}
\mathcal{L}_\chi  = \bar{\chi}\left( i \partial_\mu \gamma^\mu - g_\chi A'_\mu \gamma^\mu - m_\chi\right)\chi \; ,\label{eq:Majorana}
\end{equation}
where $g_\chi = q_{\chi_L} g' = q_{\chi_R} g'$.
Finally, the new terms involving only the dark photon are
\begin{equation}
\mathcal{L}_{A'} = -\frac{1}{4} F^{\prime \mu \nu} F^\prime_{\mu\nu} -   \frac{\epsilon}{2} F^{\mu\nu} F'_{\mu\nu}  \; ,
\end{equation}
where $F^{\prime \mu \nu}$ and $F^{\prime \mu \nu}$ denote the field strength tensors of the $U(1)'$ and $U(1)_Y$ gauge groups and $\epsilon \ll 1$ is the kinetic mixing parameter.

In summary, our model introduces six new parameters: the three masses $m_\phi$, $m_{A'}$ and $m_\chi$, the coupling $g_\chi$ and the two mixing parameters $\theta$ and $\epsilon$. 
While $m_{A'}$ and $m_\phi$ are both proportional to the vacuum expectation value of the dark Higgs field, $m_\chi$ is an independent parameter that could in principle take very different values. Here we focus on the case that all three masses are comparable in magnitude.
For simplicity, we consider three benchmark scenarios, in which the mass ratios of the three particles are fixed:
\begin{align}
&\mathbf{B1:} (\mAp, m_\chi, m_\phi)= \mAp \left(1, \frac{1}{3}, \frac{1}{6}\right), \label{def:b1}\\
&\mathbf{B2:} (\mAp, m_\chi, m_\phi)= \mAp \left(1, \frac{1}{4}, \frac{1}{6}\right), \label{def:b2}\\
&\mathbf{B3:} (\mAp, m_\chi, m_\phi)= \mAp \left(1, \frac{1}{4}, \frac{1}{8}\right) \; . \label{def:b3}
\end{align}
In each case, the chosen mass hierarchy ensures that the decays $A' \to \chi \bar{\chi}$ and $A' \to \chi \bar{\chi} \phi$ are kinematically allowed but the decay $\phi \to \chi \bar{\chi}$ is forbidden. In fact, we focus on the case that $m_\phi < m_\chi$, such that the annihilation processes $\chi \bar{\chi} \to \phi \phi$ is allowed even for vanishing initial velocities.\footnote{We note that the annihilation process $\chi\bar{\chi} \to \text{SM} \ \text{SM}$ via an off-shell dark photon is always allowed in the parameter space that we consider. However, the corresponding cross section is proportional to $\epsilon^2$ and therefore too small to deplete dark matter in the early universe in the parameter regions of interest. However, $\epsilon$ is large enough for the dark sector particles to maintain kinetic equilibrium with the Standard Model, such that all particles share a common temperature.}

\begin{figure}[t]
\centering
\includegraphics[width=0.24\linewidth]{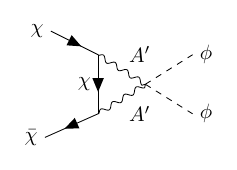}
\includegraphics[width=0.24\linewidth]{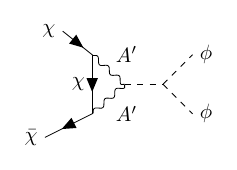}
\includegraphics[width=0.24\linewidth]{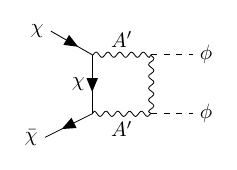}
\includegraphics[width=0.24\linewidth]{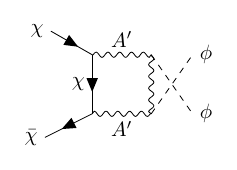}
\caption{Feynman diagrams for the dark matter annihilation process $\chi \bar{\chi} \to \phi \phi$. Since the dark matter particle does not couple to the dark Higgs boson directly, the process only happens at loop level (triangle or box).}
\label{fig:loop}
\end{figure}

In the most common version of the dark Higgs model, the dark matter particle directly couples to the dark Higgs boson, such that annihilation proceeds at tree level. For sub-GeV dark matter, however, this annihilation process would be too efficient in the early universe in the sense that it would deplete the dark matter abundance below the observed value unless the dark sector couplings are tiny (which would make them impossible to probe experimentally). Here we therefore instead consider the case that annihilation into dark Higgs bosons can only proceed via the loop-level diagrams 
shown in figure~\ref{fig:loop}.

In the non-relativistic limit, the resulting dark matter annihilation cross section takes the form
\begin{equation}
\sigma v = \frac{A \, g_\chi^8 m_\chi^2}{128 \pi^5 m_{A'}^4} v^2 \; ,
\end{equation}
where $A$ is a number of order unity that depends only on the mass ratios of the different particles, i.e.\ on the choice of benchmark scenario. For each benchmark scenario, the requirement to reproduce the observed dark matter relic abundance therefore defines a specific combination of $m_\chi$ and $g_\chi$. The relations are
\begin{align}
&\mathbf{B1:} \ m_\chi^\text{relic} = (0.89 \, \mathrm{GeV}) \, g_\chi^4, \label{eq:rel_den_b1}\\
&\mathbf{B2:} \ m_\chi^\text{relic} = (0.07 \, \mathrm{GeV}) \, g_\chi^4, \label{eq:rel_den_b2}\\
&\mathbf{B3:} \ m_\chi^\text{relic} = (0.66 \, \mathrm{GeV}) \, g_\chi^4. \label{eq:rel_den_b3}
\end{align}
The slightly smaller value in the second case is the result of a partial cancellation between diagrams with and without dark Higgs boson propagator. In all three benchmark scenarios, we find that dark matter masses around 1 GeV correspond to dark gauge couplings $g_\chi \sim \mathcal{O}(1)$.

\begin{figure}
    \centering
    \includegraphics[width=0.4\linewidth]{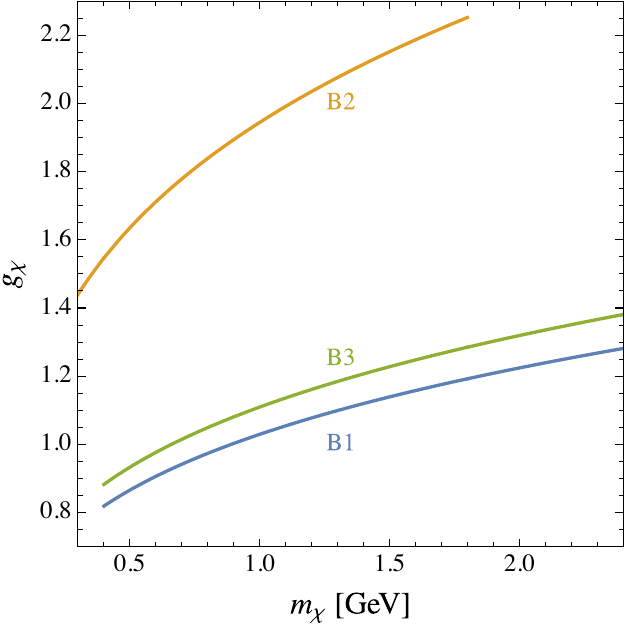}
    \caption{Value of $g_\chi$ needed to reproduce the dark matter relic density for each dark sector mass benchmark, as from \cref{eq:rel_den_b1,eq:rel_den_b2,eq:rel_den_b3}}
    \label{fig:gchi_mchi_relic_sensity}
\end{figure}

In principle, we could use these relations to fix $g_\chi$ as a function of $m_\chi$, such that the correct DM relic abundance is reproduced at every point in parameter space. However, we find that the cross section and kinematic distributions are almost completely independent of $g_\chi$, as long as $g_\chi \gg \epsilon e$, such that $\text{BR}(A' \to \chi \bar{\chi}) \approx 1$, and $g_\chi \ll 4\pi$ and the narrow-width approximation holds. We therefore take a more agnostic approach and simply fix $g_\chi = 1$ for our analysis. \Cref{fig:gchi_mchi_relic_sensity} shows that for all benchmarks this choice is not far from the $g_\chi$ that is needed to saturate the relic density. With this choice, and for a given benchmark scenario, only one mass parameter and the two mixing parameters remain free.

The kinetic mixing parameter $\epsilon$ is crucial for the production of dark sector particles from electron-positron initial states, since the direct production of dark Higgs bosons is suppressed by the tiny Yukawa coupling of the electron. For $g_\chi \gg \epsilon e$, the dark photons decay almost exclusively into dark matter particles, such that the kinetic mixing parameter is irrelevant for the dark photon branching ratios and decay length. Hence, all kinematic distributions are independent of $\epsilon$ and the total cross sections simply scale proportional to $\epsilon^2$.

For the dark Higgs boson, the situation is reversed. Its production proceeds via the dark sector gauge coupling and is independent of $\theta$. However, since only decays into SM particles are kinematically allowed, the mixing parameter affects the total dark Higgs decay width as $\Gamma_\phi \propto \theta^2$, and hence the dark Higgs boson lifetime $\tau_\phi = \Gamma_\phi^{-1}$. The total cross section is therefore independent of $\theta$, but the distribution of the dark Higgs decay vertex depends sensitively on $\theta$. In the mass range that we will be interested in, the dominant decay modes are $\phi \to \mu^+ \mu^-$ and $\phi \to \pi \pi$. We take the corresponding decay widths from Ref.~\cite{Winkler:2018qyg,Ferber:2023iso}, noting the considerable theory uncertainties~\cite{Blackstone:2024ouf}. We limit ourselves to $m_\phi < 1.2 \, \mathrm{GeV}$ so that final states with higher multiplicities, such as $\phi \to 4 \pi$, are negligible. This bound also ensures that $m_{A'} < \sqrt{s}$ for all benchmark models.

\section{Event simulations}
\label{sec:simulation}

Having discussed the model of interest, let us now turn to the simulation of signal and mock background events. In the rest of this work we will refer to the dark Higgs-strahlung process (left panel of figure~\ref{fig:fey_diags}) as the $\phi$ production process and to the radiative dark Higgs-strahlung (right panel of figure~\ref{fig:fey_diags}) as the $\gamma\phi$ production process. 

\subsection{Signal}

All signal events are generated in \texttt{MadGraph5\_aMC@NLO} \cite{madgraph}. This Monte Carlo generator weights the $i$th event by the quantity $w_i^0$ such that the cross section $\sigma^0=\sum w_i^0$. By default, the $w_i^0$ are all equal, but we will modify them as discussed below. We use the Inelastic Dark Matter UFO model that has been developed in the context of Ref.~\cite{Duerr:2020muu}, setting the inelastic mass splitting to zero to recover the case of elastic dark matter. 

We simulate events at a symmetric $e^+e^-$ collider with centre-of-mass energy $\sqrt{s}=10.58$ GeV. For the $\gamma\phi$ process, we check for the final photon to be within the Belle~II acceptance: its energy threshold is $E_\gamma \geq E_0=0.25$ GeV; its angular acceptance is $22^o \leq \theta_\gamma^{\text{cms}} \leq 180^o-22^o$~\cite{Belle-II:2010dht}. To study the dark Higgs displacement, however, we boost events to the lab frame, where the $e^-$ beam has energy $E^-=7$ GeV and tilt around the $y$ axis $\alpha_y^-=0.0415$ rad and the $e^+$ beam has energy $E^+=4$ GeV and tilt around the $y$ axis $\alpha_y^+=-0.0415$ rad.

\begin{figure}[t]
\centering
\includegraphics[width=0.45\linewidth]{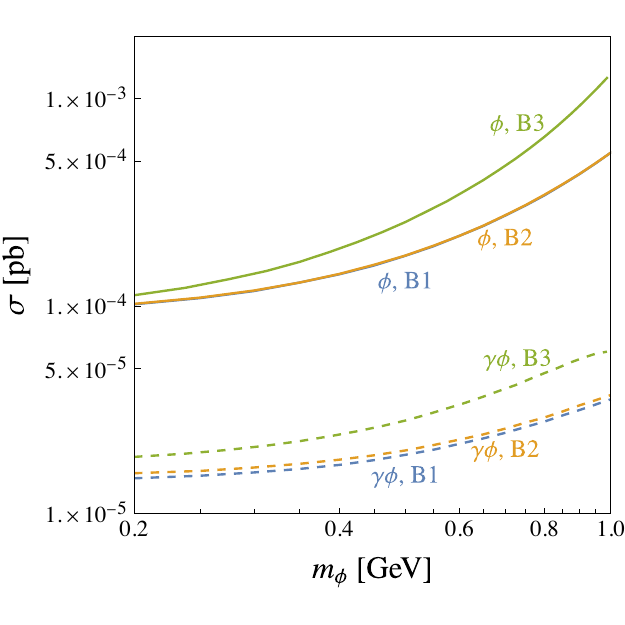}
\caption{Cross sections computed by MadGraph for the $\phi$ (solid lines) and the $\gamma \phi$ signals (dashed lines) for the benchmarks 1 (\cref{def:b1}, blue lines), 2 (\cref{def:b2}, orange lines) and 3 (\cref{def:b3}, green lines) for the benchmark value $\epsilon=10^{-4}$.}
\label{fig:xsec_plot}
\end{figure}

In figure~\ref{fig:xsec_plot}, we show the total signal cross section for kinetic mixing $\epsilon = 10^{-4}$, which is well below current bounds~\cite{BaBar:2017tiz}. As expected, the cross section for the $\phi$ process is larger than for the $\gamma \phi$ process, but the difference is only about an order of magnitude, which is much less than the factor of $\alpha$ expected for an ISR photon, because the $\gamma \phi$ process benefits from additional resonant enhancement due to the on-shell dark photon. The fact that the dark photon in the $\phi$ process must be off-shell is also the reason why the cross section grows rapidly with increasing $m_\phi$ (and therefore $m_{A'}$). But also the $\gamma \phi$ process exhibits some growth in cross section with increasing $m_\phi$ due to the soft divergence of the ISR photon.
Most importantly, however, we conclude that for both the $\phi$ and $\gamma\phi$ processes the cross sections are sufficiently high to expect a significant number of signal events at Belle~II with integrated luminosity of thousands of inverse femtobarn.

\begin{figure}[t]
\centering
\includegraphics[width=0.48\textwidth]{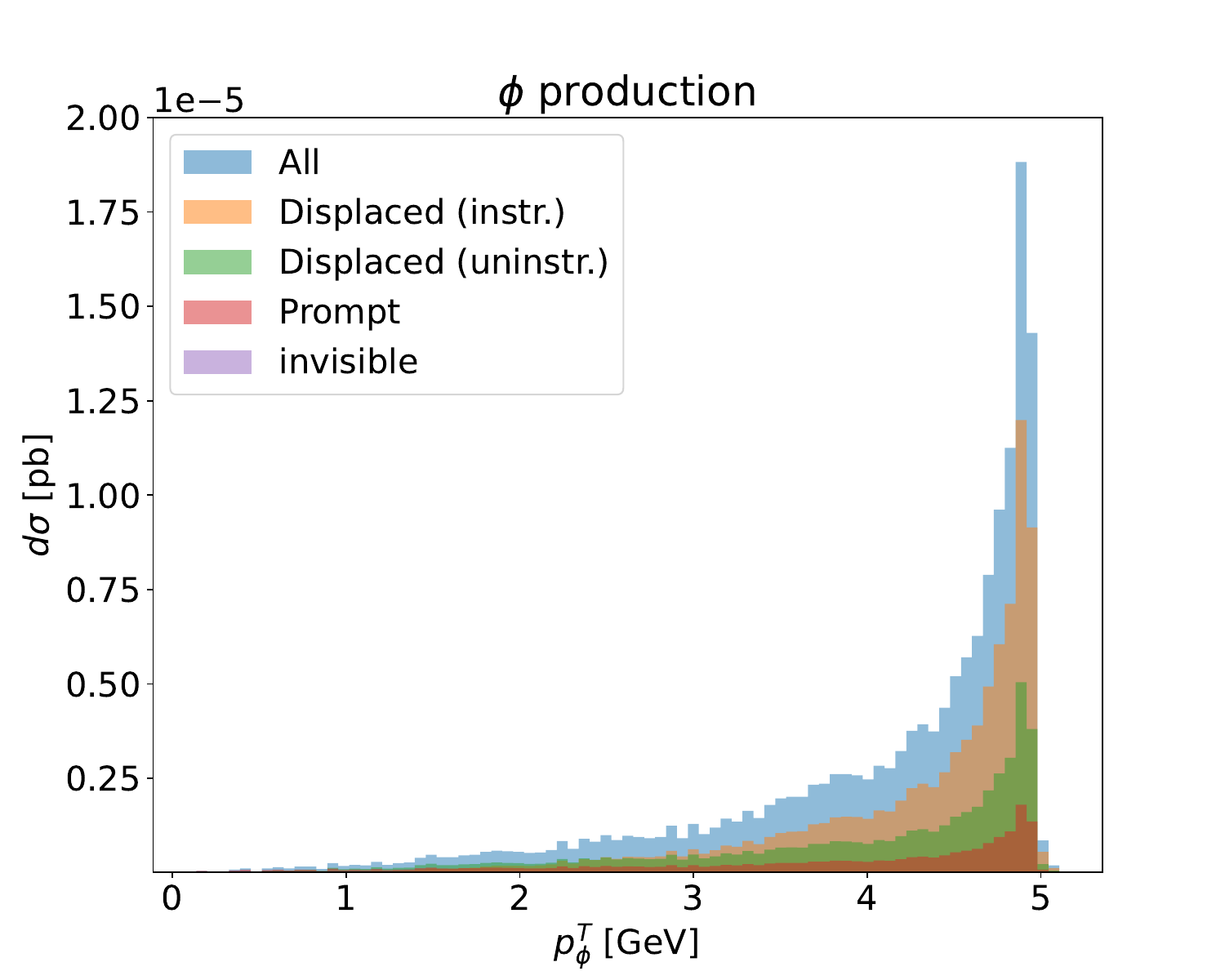}
\hfill
\includegraphics[width=0.48\textwidth]{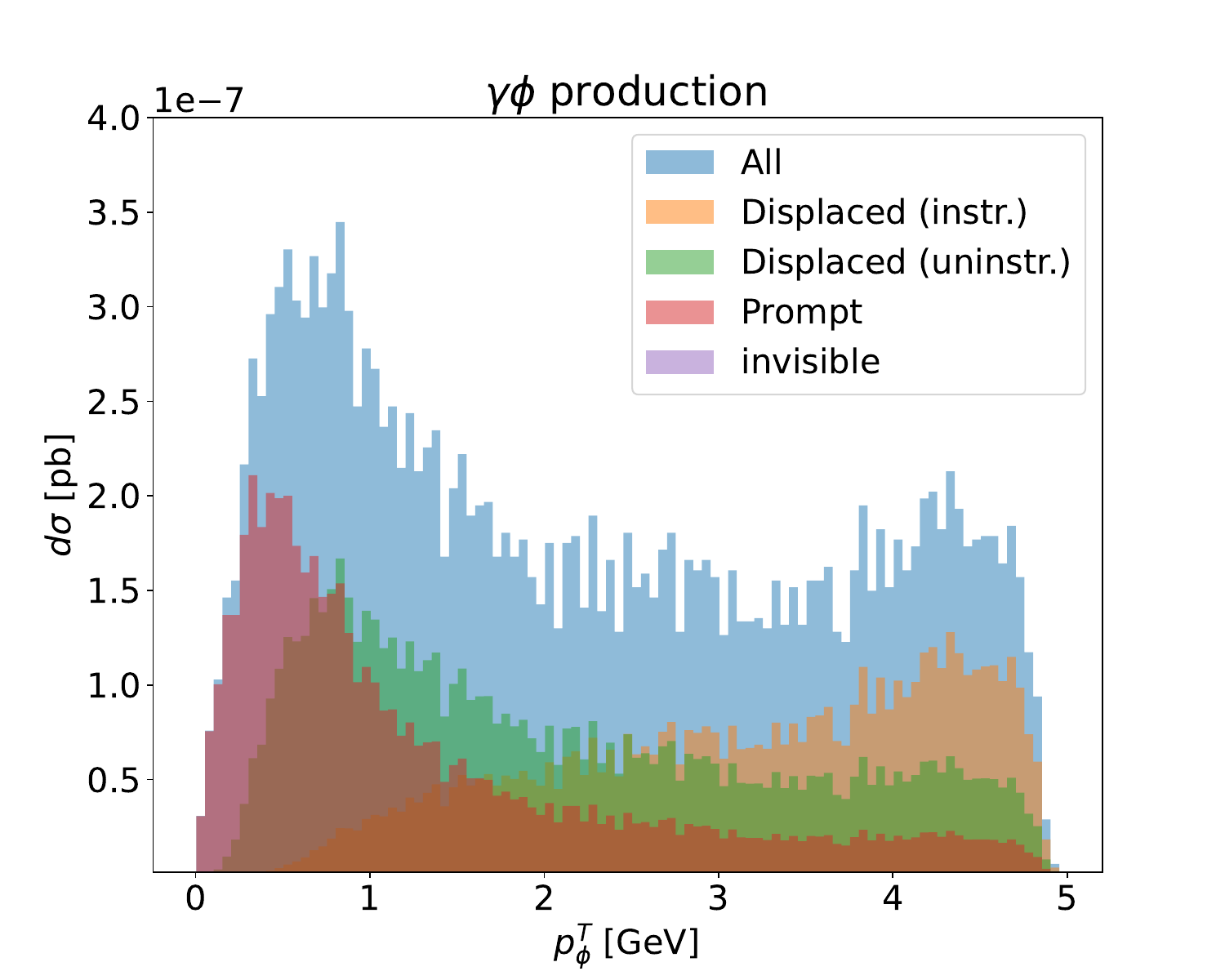}

\includegraphics[width=0.48\textwidth]{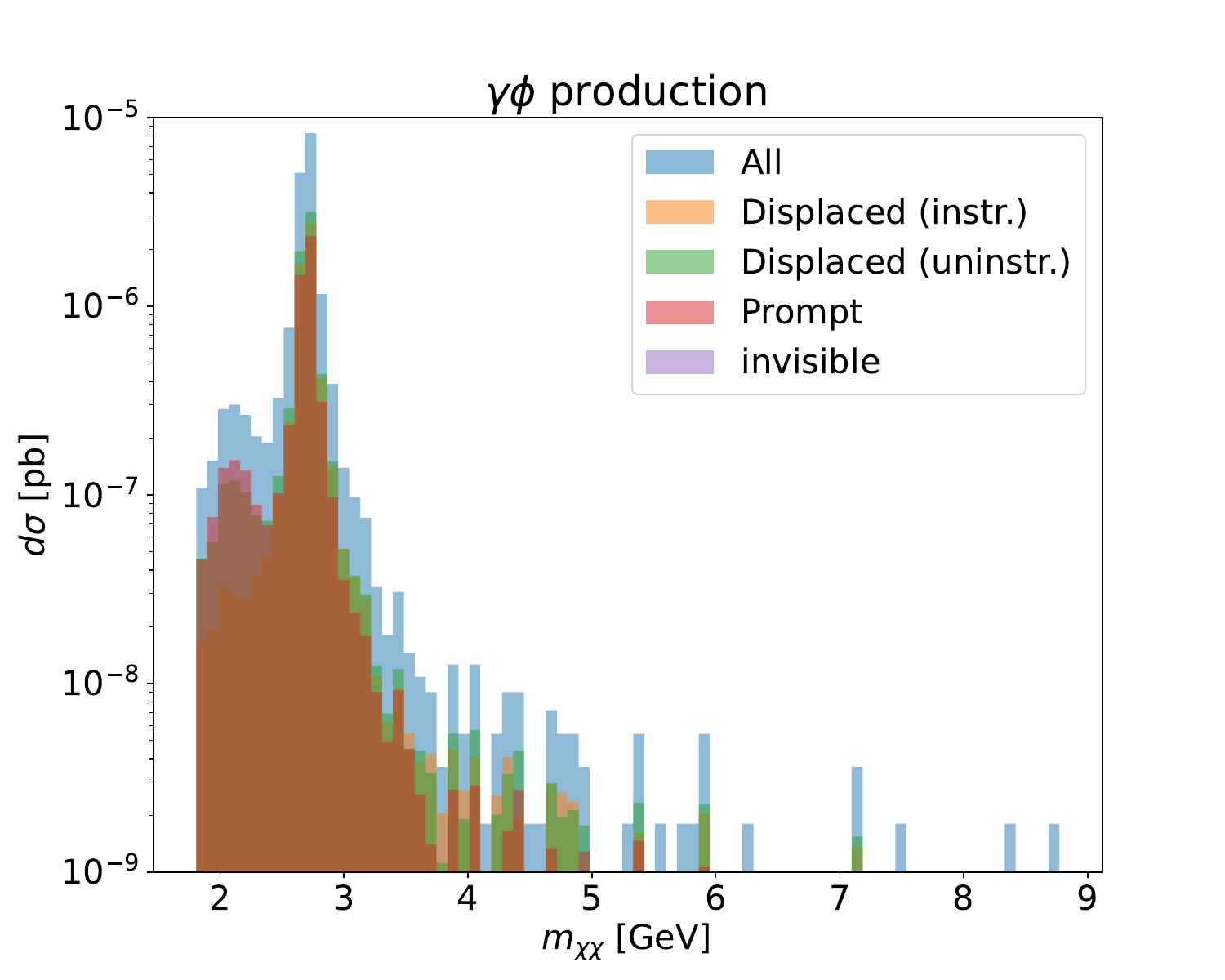}
\hfill
\includegraphics[width=0.48\textwidth]{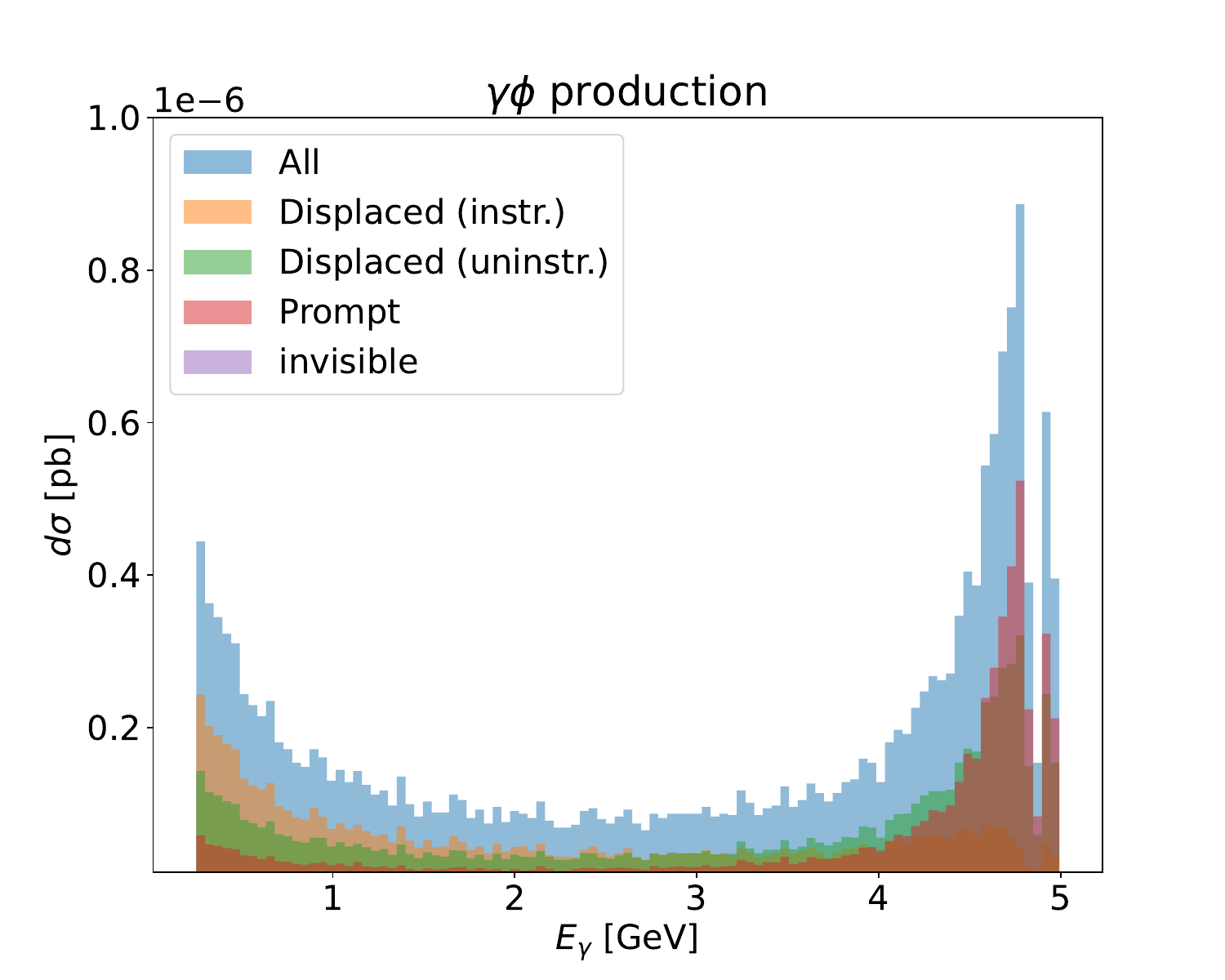}

\caption{Top row: Effect of the decay probability reweighting on cross section distribution with respect to the dark Higgs boson transverse momentum $p_\phi^T$ for $\phi$ production (left) and $\gamma \phi$ production (right). Bottom row: Effect of reweighting on other kinematic variables for $\gamma \phi$ production. The simulation parameters are $\mAp=2.7$ GeV, benchmark 1, $\theta=10^{-4}$. The legends follow \cref{tab:displacement_definitions}.}
\label{fig:dist_disp}
\end{figure}

Having generated events containing dark Higgs bosons, we need to account for the probability of the particles decaying within a given region at Belle~II. The different regions can be defined as annular cylinders with $r_\text{in} \leq \sqrt{x^2 + y^2} \leq r_\text{out}$, where the inner and outer radii are given in table~\ref{tab:displacement_definitions}. In the following, we will focus on the uninstrumented region, where photon conversion backgrounds are expected to be particularly small~\cite{Jaeckel:2023huy}.
We therefore reweight the simulated events according to the chance of a dark Higgs boson decay in this region, i.e.\ we replace each weight $w_i^0$ by a new weight $w_i \equiv p_i \cdot w_i^0$ with
\begin{align}
& p_i \equiv \exp\left(-\dfrac{r_\text{in}}{d_i}\right)-\exp\left(-\dfrac{r_\text{out}}{d_i}\right),\label{def:dec_prob}\\
& d_i \equiv \dfrac{p_i^T(\phi)}{\Gamma_\phi(\theta) m_\phi},\label{def:mean_dist}
\end{align}
where $\Gamma_\phi(\theta)$ is the full width of the dark Higgs boson as discussed in Ref.~\cite{Ferber:2023iso}. In figure~\ref{fig:dist_disp} we show the effect of this reweighting on various kinematical variables relevant for the analysis. 

\begin{table}[t]
\centering
\begin{tabular}{lcc}
\toprule
Name & $r_\text{in}$ [mm] & $r_\text{out}$ [mm]\\
\midrule
Prompt & 0 & 2\\
Displaced (uninstrumented) & 2 & 9\\
Displaced (instrumented) & 9 & $1.1 \cdot 10^3$\\
Invisible & $1.1 \cdot 10^3$ & $\infty$ \\
All & 0 & $\infty$\\
\bottomrule
\end{tabular}
\caption{To distinguish different signatures and background rates, we divide the different parts of the Belle~II detector into annular cylinders. Their axis is the beampipe, their inner radius $r_\text{in}$ and outer radius $r_\text{out}$. In our study we focus on the case that the dark Higgs bosons decay in the displaced (uninstrumented) region.}
\label{tab:displacement_definitions}
\end{table}

\subsection{Background}

The signal events that we have generated belong to two categories: one contains missing energy and a displaced vertex (with total momentum pointing back to the interaction point), the other contains an additional ISR photon. At first sight, it seems very difficult to conceive of SM backgrounds for both of these signatures. QED processes with particles missed by the detector (or weak processes involving neutrinos) are not expected to give rise to a displaced vertex. The only exception is photon conversion, which should however be small in the uninstrumented region, see the discussion in Refs.~\cite{Ferber:2022ewf,Jaeckel:2023huy}. Processes involving $K_S$ can in principle give a displaced vertex (as well as missing energy from a $K_L$), but these events are easy to identify due to the invariant mass of the decay products.

Nevertheless, in the recent search for inelastic dark matter at Belle~II~\cite{Belle-II:2025bhd} it has become clear that even these exotic signatures are not background-free. At the ultra-high luminosities and low number of signal events under consideration, very rare decay or scattering processes, misidentified  final states and  misreconstructed vertices may conspire to create background events, see Ref.~\cite{Ecker2024_1000176696} for a more detailed discussion \footnote{Exploiting the displaced vertex is a promising strategy already implemented in other BSM searches. An alternative road could be taking advantage of the polarization of Belle~II beams. Beam polarization was recently proposed in Ref.~\cite{Forti:2022mti} and used as a significance enhancing tool in BSM searches, see e.g.\ Ref.~\cite{Bauer:2023loq}. In our case for the $\phi$ search we could prepare the initial state in a spin 1. This would leave the signal unchanged with resect to unpolarized bins, while all the candidate backgrounds would be affected in total cross section and kinematic distribution. For example the background channel $e^+ e^- \gamma \gamma$ would be reduced by a factor 1/3.}. Estimating such backgrounds with theoretical tools is impossible, and deriving data-driven background estimates is clearly beyond the scope of this work. To nevertheless obtain some sensitivity estimates for dark Higgs boson searches, we therefore generate a purely phenomenological background distribution with an arbitrary normalisation, which we then vary to study its impact.

To simulate background events in an agnostic way, we proceed as follows. We consider a $2 \to 3$ or $2 \to 4$ process as background for the $\phi$ or $\gamma \phi$ process, respectively. The masses of the final state particles are respectively $(m_\phi,m_\chi,m_\chi)$ and $(0,m_\phi,m_\chi,m_\chi)$. The particle with mass $m_\phi$ plays the role of the pair of SM particles giving rise to a displaced vertex with invariant mass $m_\phi$; we assume that the vertex lies within the Belle~II uninstrumented region without applying any efficiency cut. The pair of particles each of mass $m_\chi$ are a proxy for simulating a continuum of missing 4-momentum with missing mass in the interval $[2 m_\chi,\sqrt{s}-m_\phi]$. We assume that these particles are undetected without applying any efficiency cut.

\begin{figure}[t]
\centering
\begin{subfigure}{0.48\textwidth}
\centering
\includegraphics[width=1\linewidth]{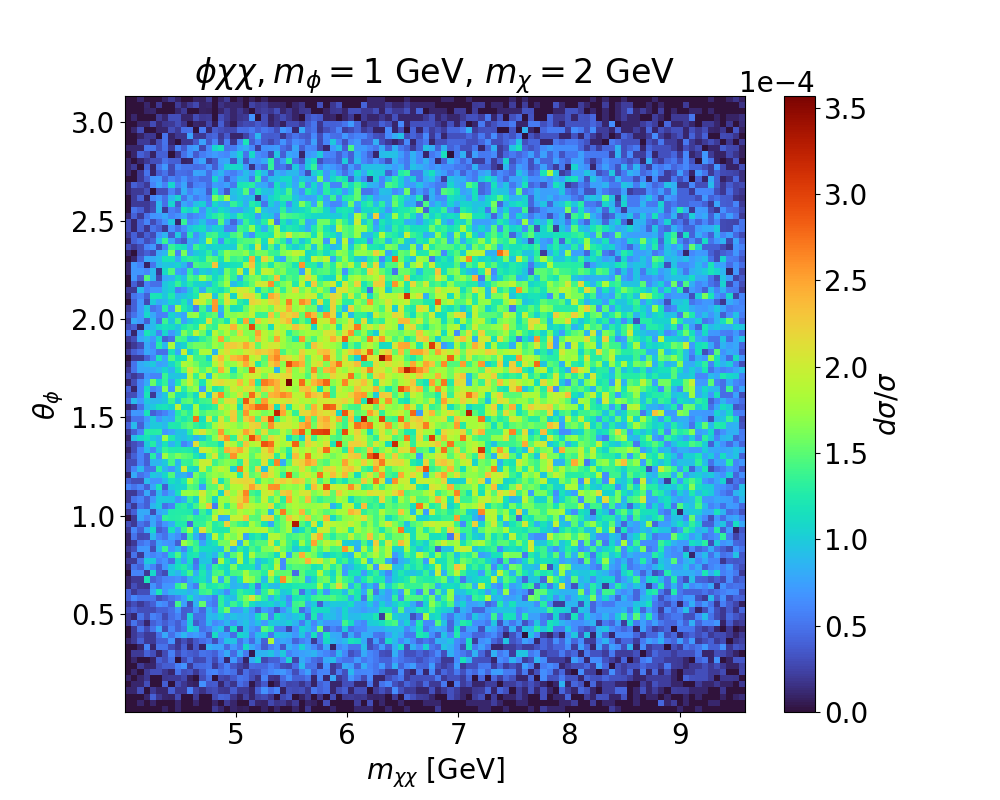}
\caption{}
\label{fig:noa_thphi_mcc}
\end{subfigure}
\begin{subfigure}{0.48\textwidth}
\includegraphics[width=1\linewidth]{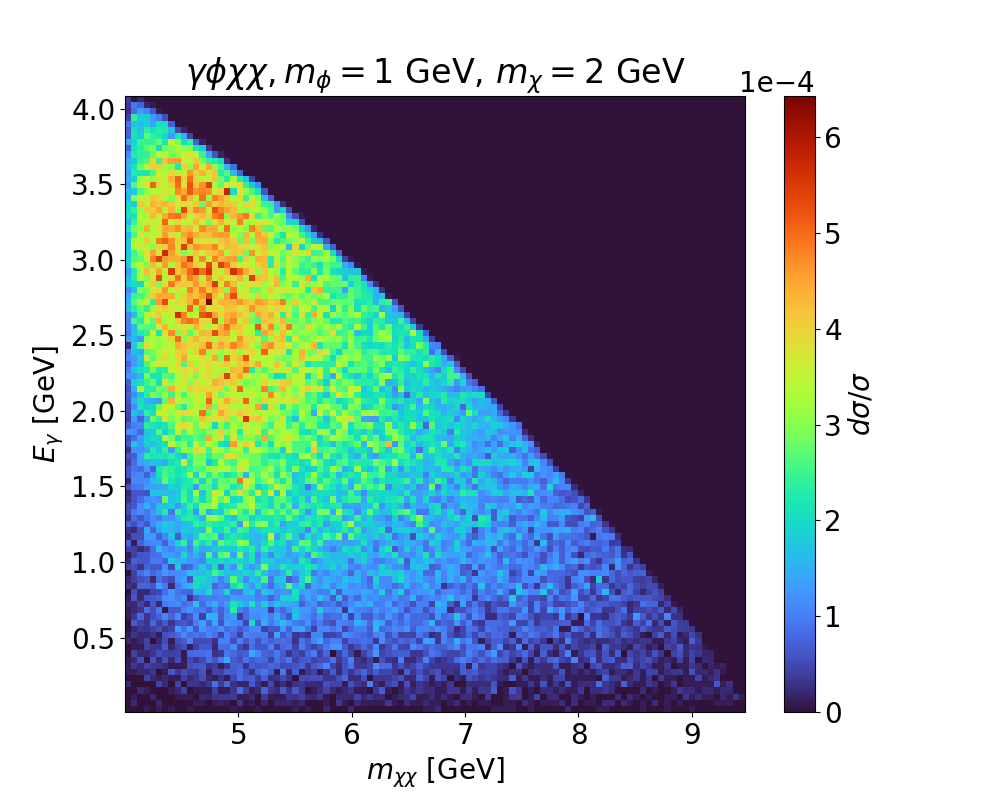}
\caption{}
\label{fig:a_ea_mcc}
\end{subfigure}
\caption{2D histograms for \texttt{RAMBO} generated mock background events. The kinematics are those of a scattering event with constant matrix element and final state particles $(\gamma)\phi \chi \bar{\chi}$. In the left panel we show the $\chi \bar{\chi}$ invariant mass against the dark Higgs polar angle for the $\phi$ production process. In the right panel we show the $\chi \bar{\chi}$ invariant mass against the dark Higgs polar angle for the $\gamma \phi$ production process. The simulation parameters are $\sqrt{s}=10.58$ GeV, $m_\phi=1$ GeV, $m_\chi=2$ GeV.}
\label{subfig:rambo_2d}
\end{figure}

Such a background simulation can be performed using the classic \texttt{RAMBO} algorithm \cite{rambo_ellis}, adapting the version coming with MadGraph \cite{madgraph} for our purposes. \texttt{RAMBO} is a Monte Carlo generator for a scattering process of $2 \to n$ particles. The input is the centre-of-mass energy and the masses of the final state particles, the output is weighted events, each described by the 4-momenta of final particles. These 4-momenta are such that the cosine of all particle polar angles are uniformly distributed, while the energies are almost uniform. If the $i^{\text{th}}$ particle has energy $q_i^0$ and three-momentum magnitude $q_i$, the distribution of $q_i^0$ is $q_i^0 e^{-q_i} \dd q_i^0$, which makes extremal energies slightly depleted of events. \texttt{RAMBO} by default returns the  integrated phase space of the $2 \to n$ process as the cross section, but we rescale the cross section such that $N_b$ background events are expected for an integrated luminosity of 50 ab$^{-1}$. We show the distributions generated by \texttt{RAMBO} in figure~\ref{subfig:rambo_2d}.

In principle, it is not clear if and how the backgrounds for the $\phi$ and $\gamma\phi$ processes are correlated. Nevertheless, we expect that, at least because of the smaller phase space, the background to $\gamma \phi$ should be smaller than the $\phi$ one. To have an estimate of the ratio between the two, we consider $e^+ e^- \to 3 \gamma$ and $e^+ e^- \to 2 \gamma$, where one of the final photons might be missed and account for missing four-momentum, and one might account for the displaced vertex through conversion. The corresponding cross sections are related by $\sigma(e^+ e^- \to 3 \gamma)=0.22 \sigma(e^+ e^- \to 2 \gamma)$. In the following, we will use this factor to set the number of expected background events for the $\gamma\phi$ process for a given number $N_b$ of background events in the $\phi$ process.

\section{Distribution of signal events}\label{sec:distributions}

The kinematics of $\phi$ and $\gamma \phi$ production is driven by the integrable divergences induced by the different propagators. Ignoring numerators, the matrix elements of the two processes have the form
\begin{align}
& {\cal M}(\phi) \sim \dfrac{1}{s-\mAp^2+i \mAp \Gamma_{A^\prime}} \cdot \dfrac{1}{\mcc^2-\mAp^2+i \mAp \Gamma_{A^\prime}},\label{eq:mat_elem_no_a}\\
& {\cal M}(\gamma \phi) \sim \dfrac{1}{\mcc^2-\mAp^2+i \mAp \Gamma_{A^\prime}} \cdot \dfrac{1}{(p_1-p_\gamma)^2-m_e^2} \cdot \dfrac{1}{(p_1+p_2-p_\gamma)^2-\mAp^2+ i \mAp \Gamma_{A^\prime}}.
\end{align}

For $\phi$ production, the $A^\prime$ propagator leads to a peak at $\mcc=\mAp$. The signature with a photon, on the other hand, has more complicated kinematic features, stemming from the different propagators:
\begin{enumerate}
\item The electron propagator from photon radiation. Using \begin{equation}
(p_1-p_\gamma)^2-m_e^2 \approx -\sqrt{s} E_\gamma(1- \cos \theta_\gamma) \; ,
\end{equation}
we find the well-known soft divergence ($E_\gamma=0$) and collinear divergence ($\theta_\gamma=0$). Both of these are however incompatible with our requirements on the minimal photon energy ($E_\gamma > E_0$) and photon direction and can therefore not be attained exactly. 
\item The first dark photon propagator. Neglecting $\Gamma$, this propagator diverges for
\begin{equation}
E_\gamma=\bar{E}\equiv\dfrac{s-\mAp^2}{2 \sqrt{s}} \; .\label{eq:eb_sol}
\end{equation}
\item The second dark photon propagator. Just like for $\phi$ production, it diverges at $\mcc=\mAp$.
\end{enumerate}
The preference for small $\theta_\gamma$ from the electron propagator is compatible with all other requirements, such that we expect the photon angular distribution to peak close to the cut-off angle $\theta_\text{min} = 22^\circ$. However, the same is expected to be true for many background processes (which may be even more strongly peaked~\cite{b2_phys_book}), such that we do not expect the photon angular distribution to help with signal-background discrimination and will not consider it in the following.

Conversely, the distribution of $m_{\chi\chi}$ and $E_\gamma$ are expected to take a very different shape for signal and background. The conditions $\mcc=\mAp$ and \cref{eq:eb_sol} exclude one another because they correspond respectively to the first or the second virtual dark photon being on-shell. If the first dark photon is on-shell, $\mcc$ will not be strongly peaked anywhere, but it must satisfy the kinematic requirement that $\mcc < m_{A'} - m_\phi$. If, on the other hand, the second dark photon is on-shell, the photon energy is not peaked but limited by the bound 
\begin{equation}
E_\gamma < E_b \equiv\dfrac{s-(\mAp+m_\phi)^2}{2 \sqrt{s}}. \label{eq:high_ea_mcc_map}
\end{equation}
We note that for small $m_\phi$ the two energies $\bar{E}$ and $E_b$ defined respectively in eqs.~\eqref{eq:eb_sol} and \eqref{eq:high_ea_mcc_map} become very close, such that $E_\gamma$ is expected to peak at the highest energies kinematically allowed. For large $m_\phi$, instead, it may be favourable for the photon energy to be as close as possible to the lower bound $E_0$, in order to maximise the contribution from the electron propagator. In conclusion, three possible photon energy regimes can be realized:
\begin{description}
\item[Soft regime:] The photon energy peaks at $E_\gamma \sim E_0$.
\item[Hard regime:] The photon energy peaks at $E_\gamma \sim \bar{E}$.
\item[Bimodal regime:] The photon energy has peaks of similar intensity at $E_\gamma \sim E_0$ and $E_\gamma \sim \bar{E}$ (or is nearly flat).
\end{description}

\begin{figure}[t]
\centering
\begin{subfigure}{0.32\textwidth}
\includegraphics[width=\textwidth]{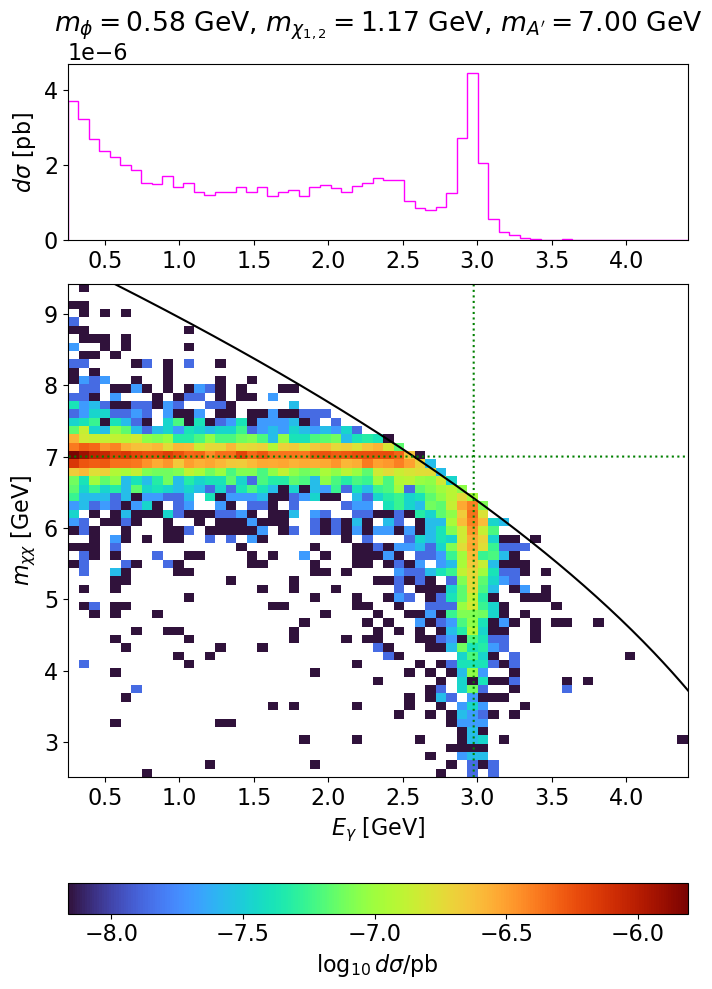}
\caption{Soft regime.}
\label{fig:ea_mcc_soft}
\end{subfigure}
\hfill
\begin{subfigure}{0.32\textwidth}
\includegraphics[width=\textwidth]{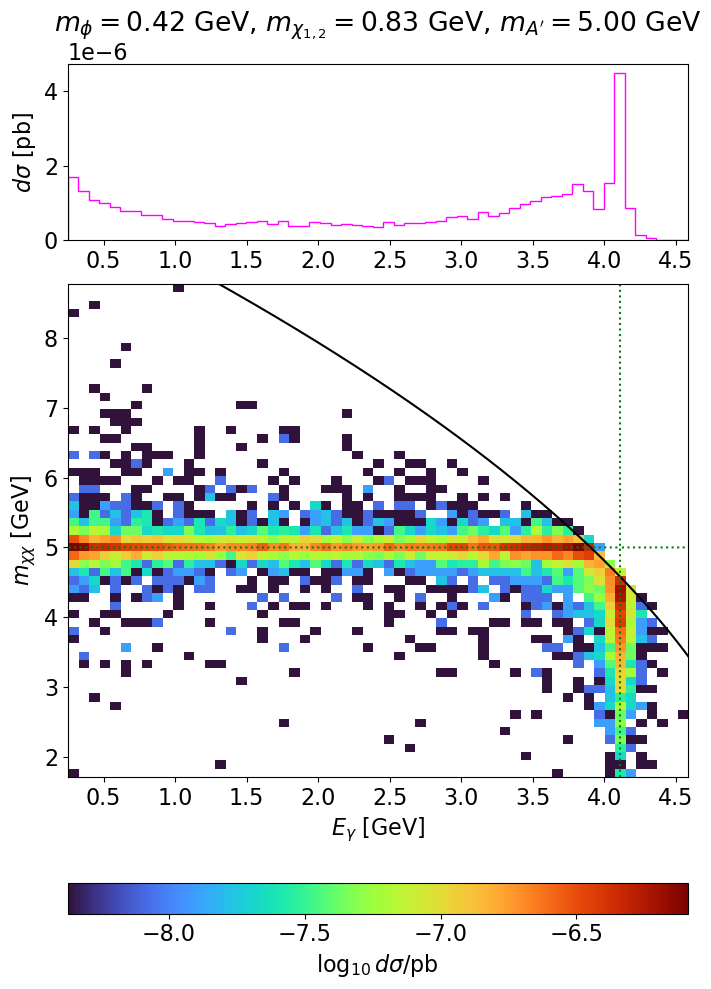}
\caption{Bimodal regime.}
\label{fig:ea_mcc_bimodal}
\end{subfigure}
\hfill
\begin{subfigure}{0.32\textwidth}
\includegraphics[width=\textwidth]{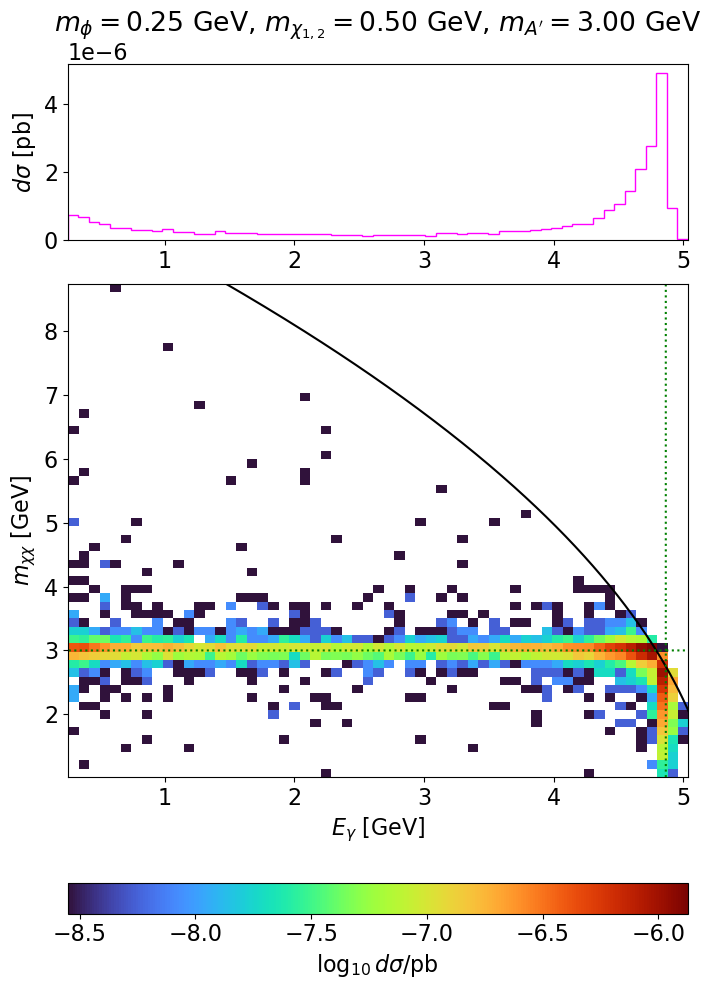}
\caption{Hard regime.}
\label{fig:ea_mcc_hard}
\end{subfigure}
\caption{Dalitz plots for $\gamma \phi$ production in the $(E_\gamma, \mcc)$ plane topped by a 1d histogram in the $E_\gamma$ variable. Both $\mcc=\mAp$ and $E_\gamma=\bar{E}$ resonances are visible ($\bar{E}$ is defined in \cref{eq:eb_sol}). The horizontal branch can be peaked at soft photons (left panel), hard photons (right), or favour neither of the two (called bimodal regime and shown in the central panel). The black solid line delimits the phase space boundary as from \cref{eq:high_ea_mcc_map}. The dotted green line is the locus of point $(\mcc-\mAp) \cdot (E_\gamma-\bar{E})=0$. In each simulation we set $\epsilon=10^{-4}$.}
\label{fig:ea_mcc_2_d_histos}
\end{figure}

The different regimes are illustrated in \cref{fig:ea_mcc_2_d_histos}. In each panel, one can identify two different branches, corresponding to the case where either the first or the second dark photon is on-shell. The distribution of events along each branch, as well as the overall weight of the two branches, depends on the relative importance of the different divergences and differs across the three regimes.
\Cref{tab:divergences} summarizes all divergences that can happen at the same time, and which kinematic configuration they correspond to.

\begin{table}[t]
\centering
\begin{tabular}{cccccc}
\toprule
$\theta_\gamma \to 0$ &  $E_\gamma \to 0$ & $E_\gamma \to \tfrac{s-\mAp^2}{2 \sqrt{s}}$ & $E_\gamma \to \tfrac{s-(\mcc +m_\phi)^2}{2 \sqrt{s}}$  & $\mcc \to \mAp$  & Branch \\
\midrule
\checkmark & \checkmark  &  &  & \checkmark & Horizontal (soft $\gamma$)\\
\checkmark &  &  & \checkmark & \checkmark & Horizontal (hard $\gamma$)\\
\checkmark &  & \checkmark &  &  & Vertical\\
\bottomrule
\end{tabular}
\caption{Table of possible divergences of $\gamma \phi$ production. The corresponding distributions are depicted in  \cref{fig:ea_mcc_2_d_histos}.}
\label{tab:divergences}
\end{table}

\begin{figure}[t]
\centering
\includegraphics[width=0.7\textwidth]{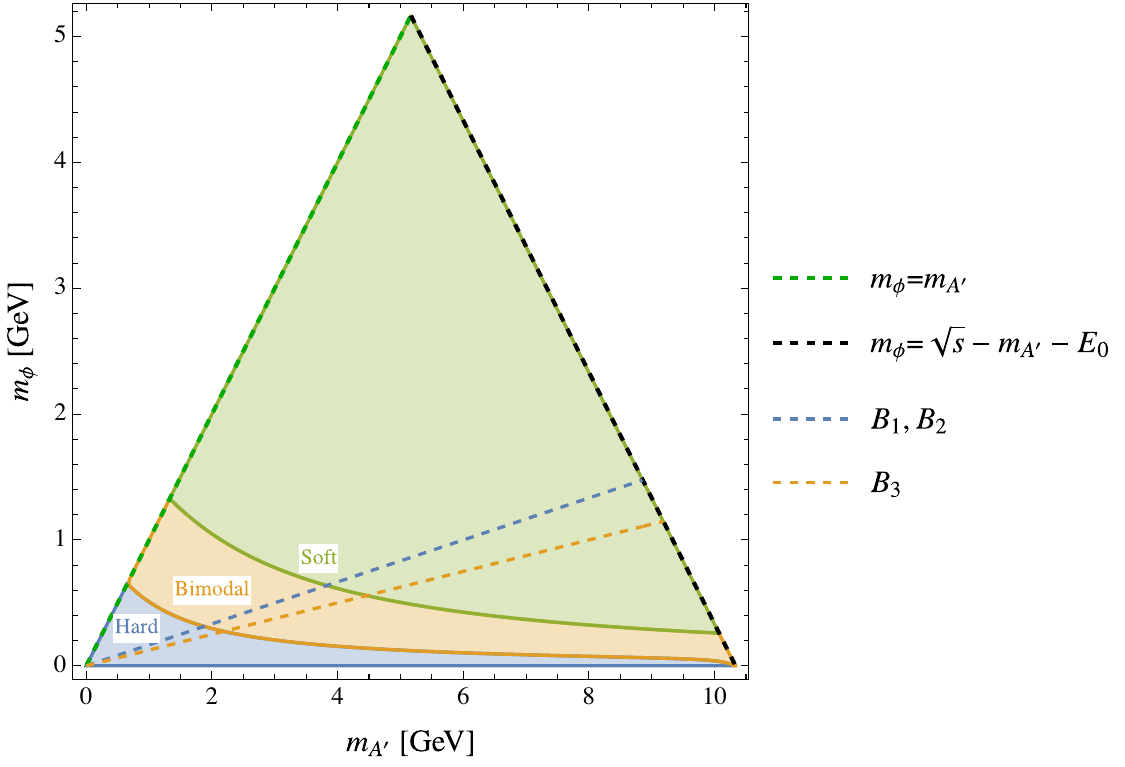}
\caption{Division of the allowed phase space for $\mAp$ and $m_\phi$ in the $\gamma \phi$ process based on the importance of the different matrix element divergences. The phase space is bounded by $m_\phi=0$ (blue), $m_\phi=\mAp$ (dashed dark green), $m_\phi= \sqrt{s}-\mAp-E_0$ (dashed black). The soft regime (green area) corresponds to the first line in \cref{tab:divergences} and prefers $E_\gamma \sim 0$ such that the $r$ defined in \cref{def:hard_soft_ratio} is small; the hard regime (blue area) corresponds to the second line in \cref{tab:divergences} and prefers $E_\gamma \sim \bar{E}$ such that the $r$ defined in \cref{def:hard_soft_ratio} is large; the bimodal regime (orange area) lies between the previous two. The dashed blue and orange lines indicate the combination of $m_{A'}$ and $m_\phi$ corresponding to our benchmark models.}
\label{fig:hbs_region_plot}
\end{figure}

Clearly, a peak in the $E_\gamma$ distribution is advantageous for the $\gamma \phi$ analysis, in particular if the peak is at high energies, giving a clean experimental signature. For this reason it is important to understand which of the regimes identified above corresponds to which combinations of $\mAp,m_\phi,m_\chi$. The propagators explicitly depend on $\mAp$ and $m_\phi$, whereas the dependence on $m_\chi$ enters implicitly through the phase space. Let us define the ratio
\begin{equation}
r \equiv \dfrac{{\cal M}(\text{hard} \gamma)}{{\cal M}(\text{soft} \gamma)}\label{def:hard_soft_ratio} \; ,
\end{equation}
which quantifies the relative importance of hard and soft photons. We can solve this equation for $m_\phi$, finding
\begin{equation}
m_\phi=\frac{1}{\sqrt{2}} \left[s+\mAp^2-\sqrt{8 E_0 \sqrt{s} \left(2 E_0 \sqrt{s}+\mAp^2-s\right)/r+(s-\mAp^2 )^2}\right]^{1/2}- \mAp.\label{eq:r01_sol}
\end{equation}
We can now define the three different regimes in terms of the ratio $r$ and hence a range of $m_\phi$:
\begin{description}
\item[Soft regime:] $r < 1$; 
\item[Bimodal regime:] $1 \leq r \leq 4$;
\item[Hard regime:] $r > 4$.
\end{description}
We depict these three regimes in terms of $m_{A'}$ and $m_\phi$ in \cref{fig:hbs_region_plot}.

\subsection{Suitable kinematic variables}

\begin{figure}[t]
\centering
\includegraphics[width=0.48\textwidth]{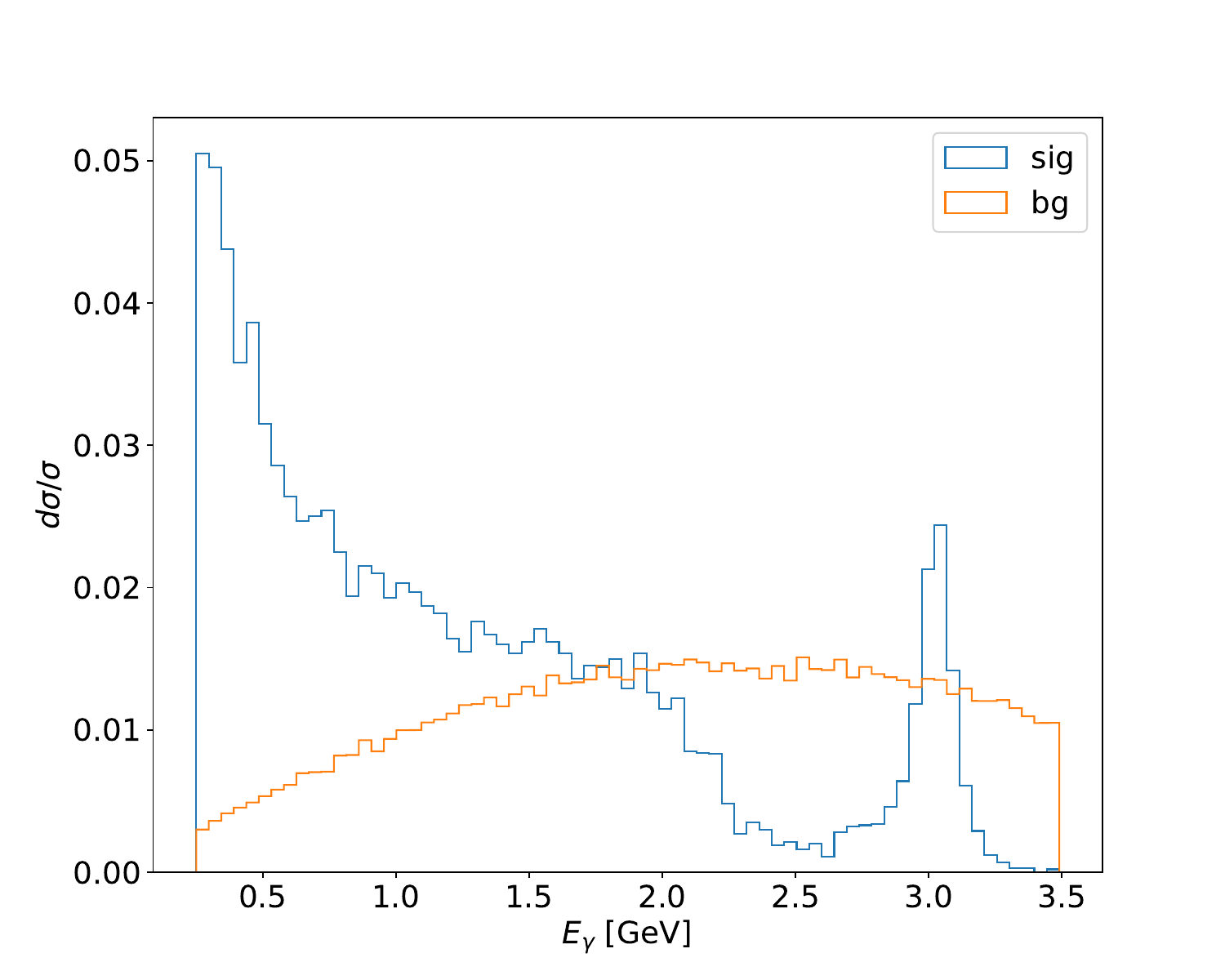}
\hfill
\includegraphics[width=0.48\textwidth]{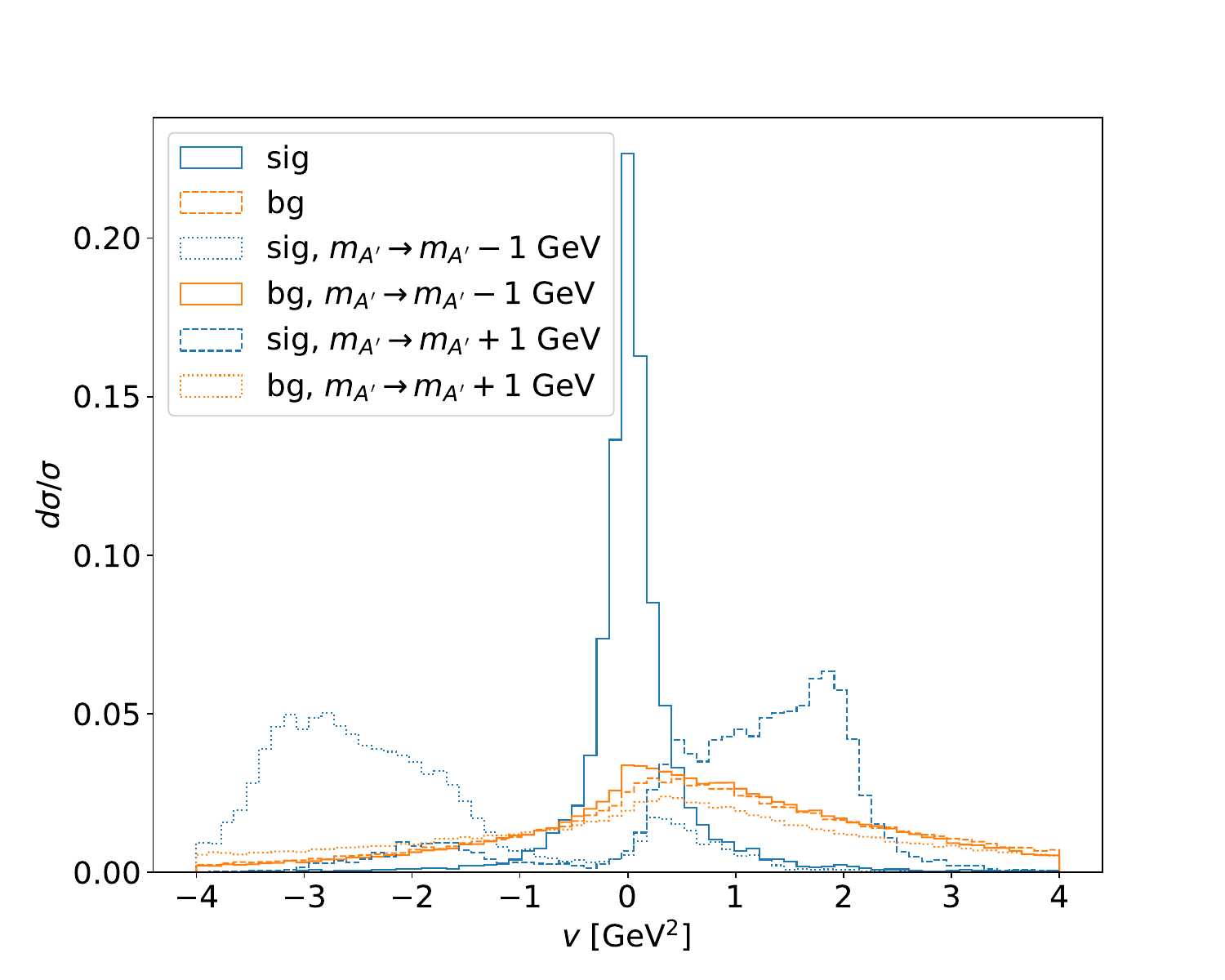}
\caption{Left: signal (blue) and background (orange) distributions with respect to the photon energy $E_\gamma$. Right: distributions with respect to the $v$ variable defined in \cref{def:v_hyperbole} for the signal onto the DP nominal mass $\mAp$ (solid blue) or onto incorrect masses $\mAp-1$ GeV (dotted blue), $\mAp+1$ GeV (dashed blue), and for the background onto $\mAp$ (solid orange), $\mAp-1$ GeV (dotted orange), $\mAp+1$ GeV (dashed orange). In both panels $\mAp=7$ and benchmark=2.}
\label{fig:ea_v_sig_bg_histos}
\end{figure}

Figure~\ref{fig:ea_mcc_2_d_histos} shows clearly that for the case of $\gamma \phi$ production it is not enough to carry out a bump hunt in either $m_{\chi\chi}$ or $E_\gamma$, because these kinematic variables have broad (or even multiple) peaks. One possibility would be to instead carry out a two-dimensional likelihood analysis in both $\mcc$ and $E_\gamma$. However, it turns out that almost the same sensitivity can be achieved using a single variable that condenses the information from both $\mcc$ and $E_\gamma$. Such a variable is
\begin{equation}
v:=\left(E_\gamma - \Eb\right) \cdot (\mcc - \mAp)\;. \label{def:v_hyperbole}
\end{equation}
The points corresponding to $v=0$ are depicted in \cref{fig:ea_mcc_2_d_histos} as green dotted lines. In \cref{fig:ea_v_sig_bg_histos} we compare signal and background distributions with respect to the photon energy (left panel) or the variable $v$ from \cref{def:v_hyperbole} (right). While in the left panel both distributions are rather broad, transforming $(E_\gamma, \mcc)$ to $v$ in the right panel causes all signal data to converge around $v\sim 0$ while the background distribution does not exhibit a strong peak. Clearly, the usefulness of $v$ stems from its narrow width. Indeed, starting from the dark photons propagators, it can be shown that the width of $v$ is ${\cal O}(\sigma_{E_\gamma} \cdot \sigma_{m_{\chi\chi}})$, where $\sigma$ denotes respectively the width of the $E_\gamma$ distribution around $\bar{E}$ or the width of the $\mcc$ distribution around $\mAp$. We expect both of these widths to be of the order $\sigma \sim \max(\Gamma_{A'}, \delta E)$ with the energy resolution $\delta E$ defined in \cref{app:smear_bin}.

We emphasize that the construction of $v$ requires a choice of $m_{A'}$. In other words, $v$ peaks at zero for the signal only if the correct dark photon mass is used.  For different values of $m_{A'}$, the distribution of $v$ is broader and peaks away from zero (see the dashed and dotted lines in \cref{fig:ea_v_sig_bg_histos}). This means that an analysis based on the variable $v$ still requires a scan over $m_{A'}$ in the same way as a bump hunt in $E_\gamma$. The only difference is that the signal would always peak in the same place ($v = 0$), but the background distribution changes slightly as $m_{A'}$ is varied.

\subsection{Event selection}

We analyze the $\phi$ channel as follows: After generating events for each mass point $(\mAp,m_\chi,m_\phi)$, we loop over possible values of the Higgs mixing $\theta$ and perform the decay probability rescaling. On the rescaled data, we consider the missing invariant mass and add a smearing effect as detailed in \cref{subsec:smearing}. We then loop over possible kinetic mixing values and rescale the signal according to $\sigma \propto \epsilon^2$. Finally, we vary the expected number of background events $N_b$ in the range [1,10$^4$] by rescaling the \texttt{RAMBO} cross section. This procedure yields histograms of $\mcc$ for signal and background for the values $(\mAp,m_\chi,m_\phi,\epsilon,\theta,N_b)$. Assuming an experimental null result, the log-likelihood of a given signal hypothesis is
\begin{equation}
\log {\cal L} = -S+B \log \left(1+ \dfrac{S}{B}\right) \, ,\label{eq:likelihood}
\end{equation}
with $S(B)$ the number of expected signal (background) events at Belle~II after cuts. We determine $\mcc^\text{max}, \mcc^\text{min}$ such that the selection $\mcc^\text{min} \leq \mcc \leq  \mcc^\text{max}$ maximises \cref{eq:likelihood}.
Concretely, we start from the bin centred around $\mAp$ and include more and more neighbouring bins until \cref{eq:likelihood} is maximized. The corresponding signal efficiencies are tabulated in \cref{app:efficiencies}. The expected Belle~II sensitivity at 95\% confidence level is then given by $-2 \log \mathcal{L} = 3.84$.

The analysis for the $\gamma \phi$ channel is very similar, except the following details: $\sqrt{|v|}$ is used\footnote{The variable $\sqrt{|v|}$ is more convenient than $v$ because its behaviour under smearing of $E_\gamma$ and $\mcc$ is more stable, see \cref{subsec:binning}.
} instead of $\mcc$; both $E_\gamma$ and $\mcc$ are smeared; the optimization algorithm is such that we include signal and background with $0 \leq \sqrt{|v|} \leq \sqrt{|v^\text{max}|}$; lastly, the binning is
\begin{equation}
e_0=0 \, , \qquad e_1=0.01 \, \GeV \, , \qquad \dfrac{e_{i+1}-e_i}{\frac{1}{2}(e_{i+1}+e_i)}=30\% \; ,\label{eq:v_bin}
\end{equation}
where $e_{i+1}$ is the right edge of the $i$th bin. The total number of bins is chosen in such a way that all signal and background events are contained in the histogram.

\begin{figure}[t]
\centering
\includegraphics[width=0.48\linewidth]{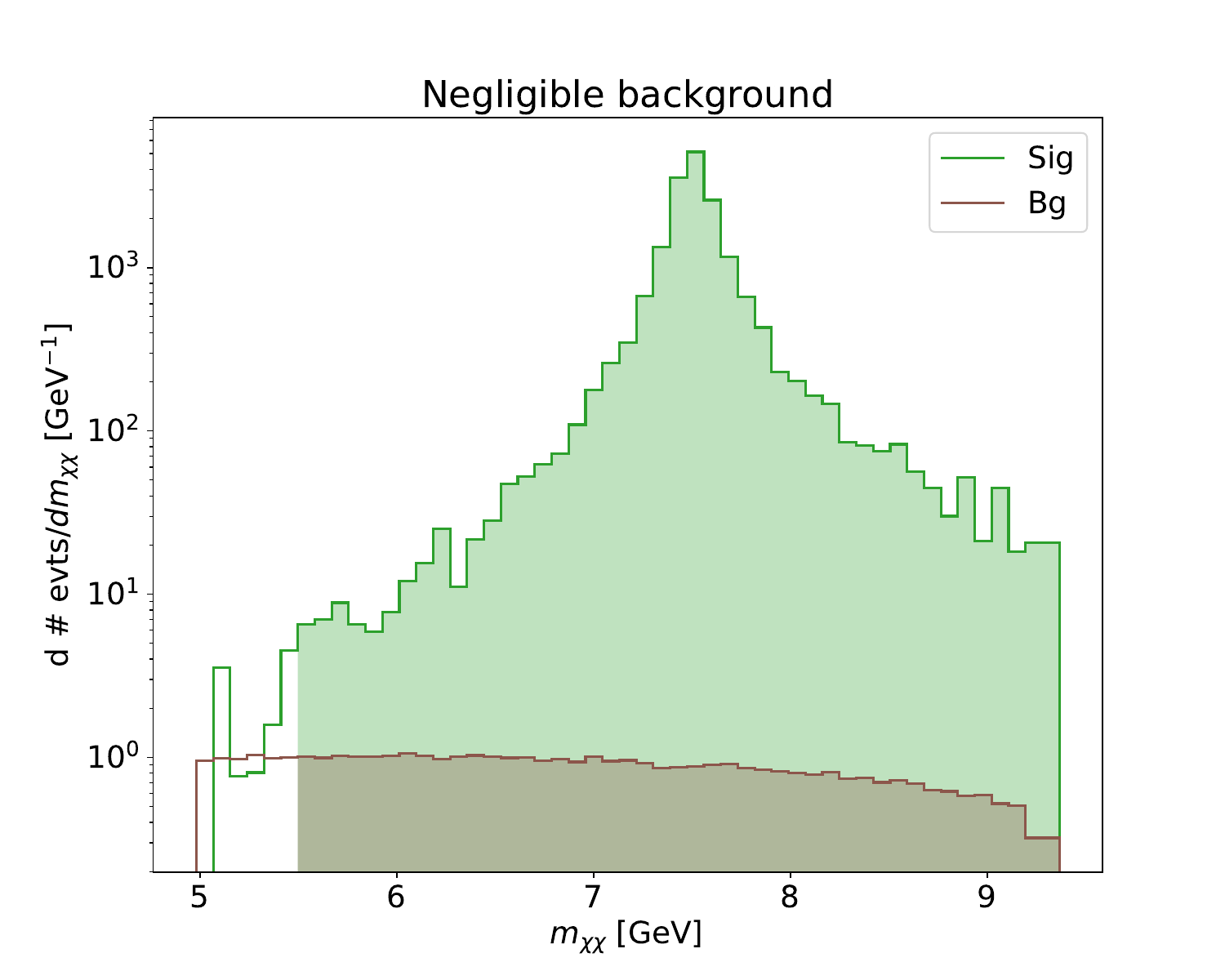}
\hfill
\includegraphics[width=0.48\linewidth]{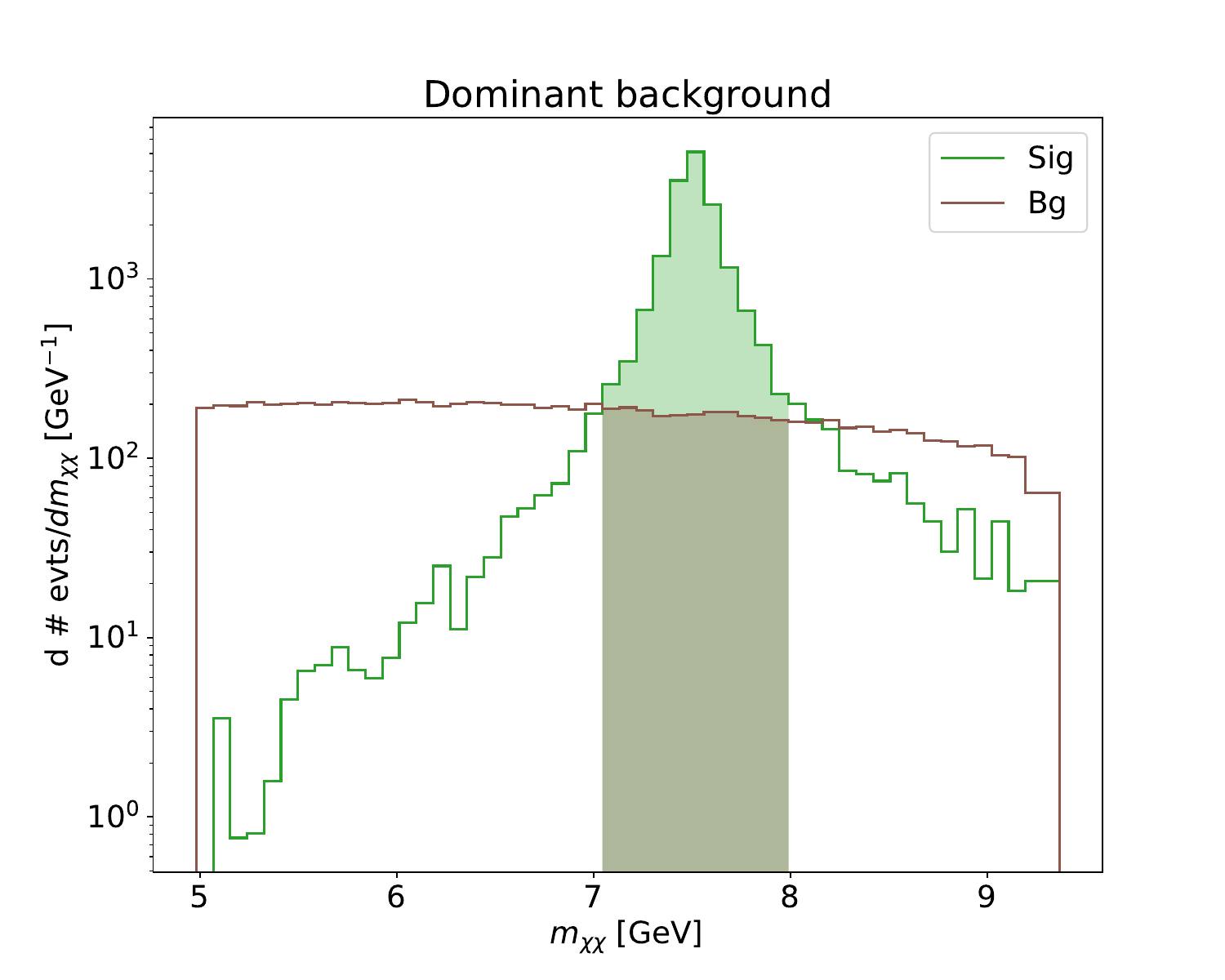}

\includegraphics[width=0.48\linewidth]{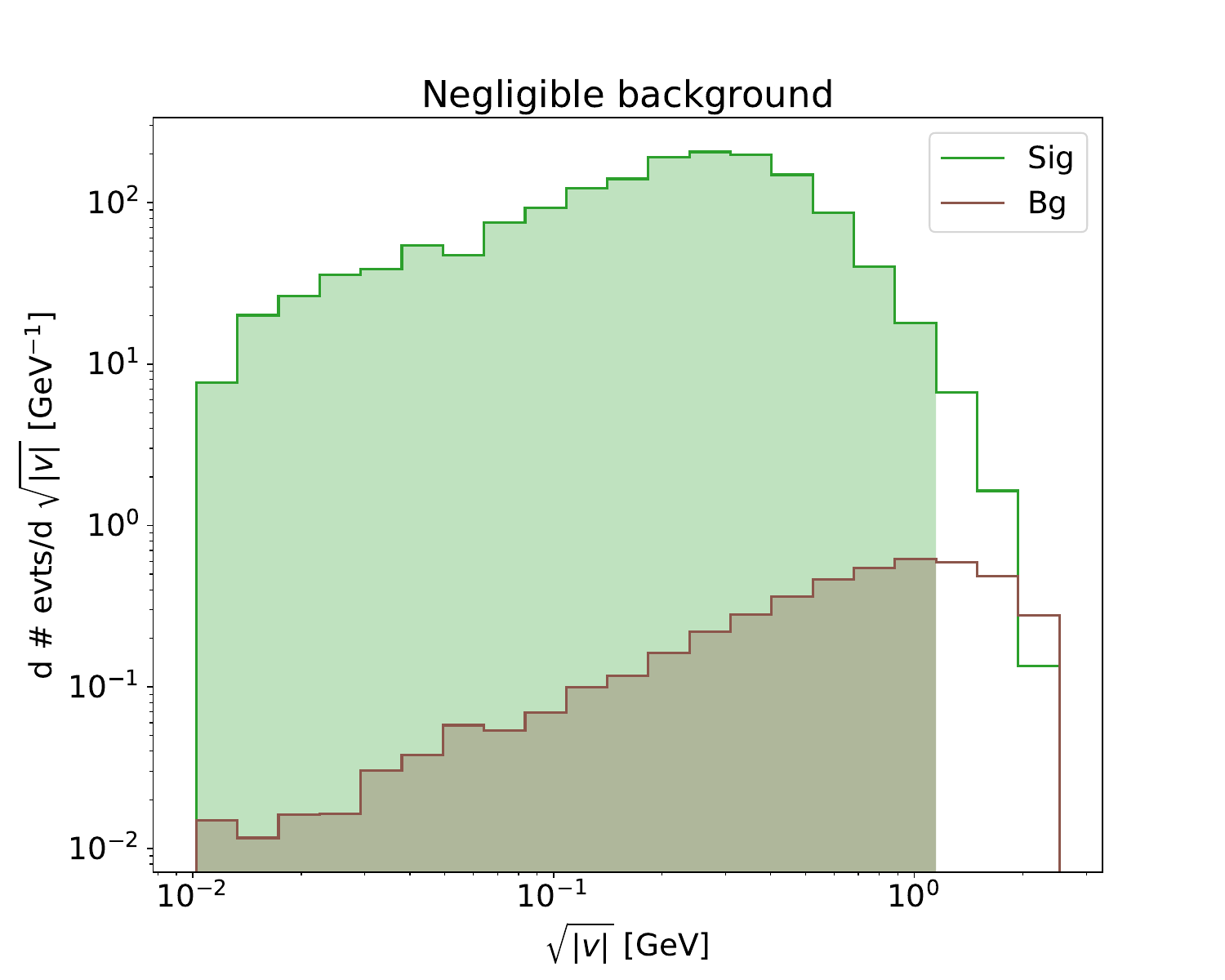}
\includegraphics[width=0.48\linewidth]{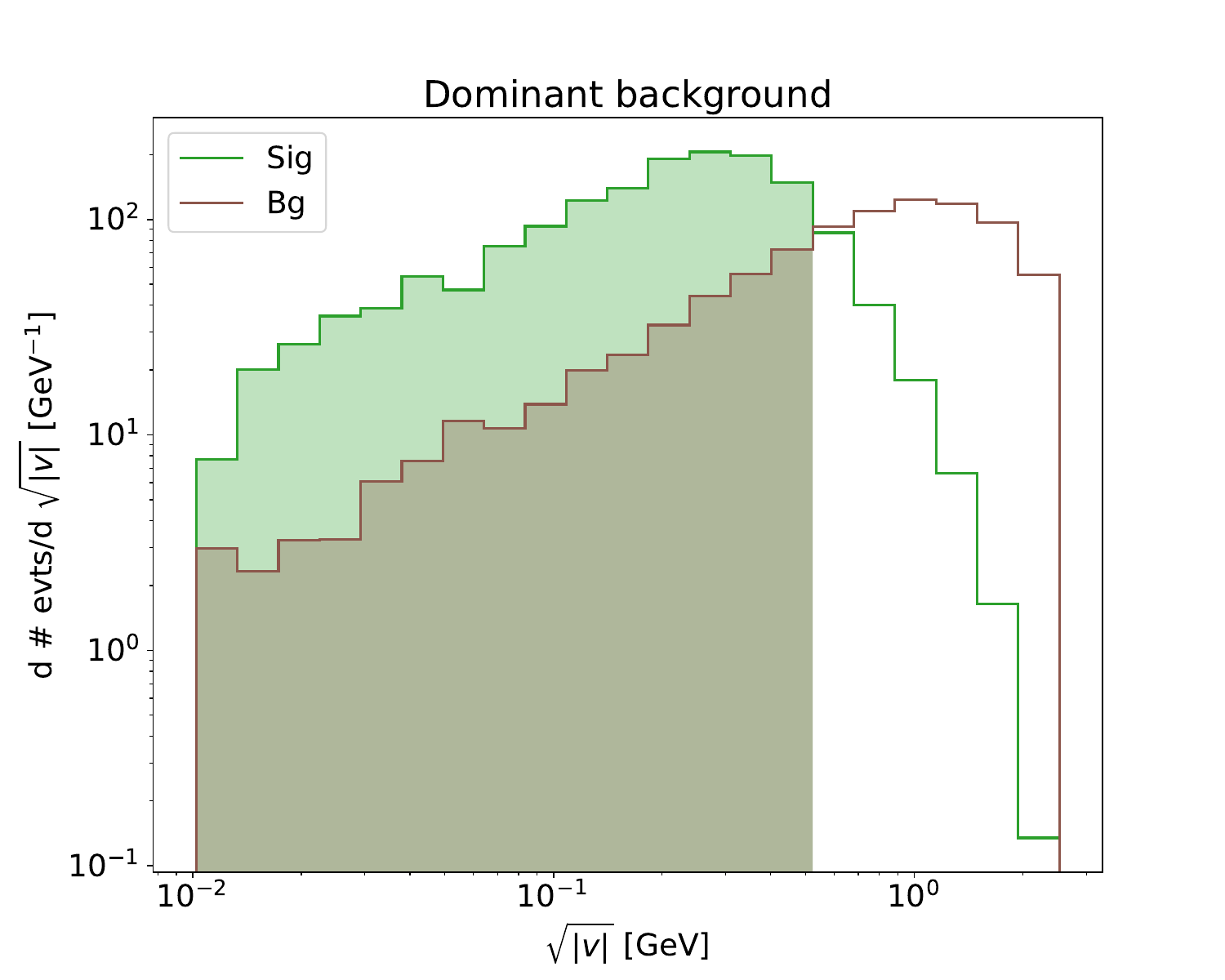}
\caption{Distributions of expected events at Belle~II for $\phi$ (top row) or $\gamma \phi$ events (bottom row) with respect to the corresponding analysis variable ($\mcc$ for the top row and $\sqrt{|v|}$ for the bottom row). Signal (background) events are represented with empty green (brown) histograms. Shaded bins are those selected by the optimisation analysis. In the left column the number of background events is negligible ($N_b=5$ for the $\phi$ process, for the $\gamma\phi$ process it is rescaled as discussed in \cref{sec:simulation}); in the right column the background is dominant ($N_b=10^3$ for the $\phi$ process). The simulation parameters are: $\mAp=7.5$ GeV, benchmark 1, $\epsilon=10^{-4}$ and $ \theta=10^{-4}$.}
\label{fig:optim_algo_histos}
\end{figure}

An example of the chosen binnings and of the optimizing algorithm output for both the $(\gamma)\phi$ processes is given in \cref{fig:optim_algo_histos}. We remark how, for negligible background, the optimizing strategy is keeping a large part of the signal while for dominant backgrounds only the peaking values $\mcc \sim \mAp$ and $v \sim0$ are selected in order to drastically reduce the background without too much harm to the signal.

\section{Results and discussion}\label{sec:results}

\begin{figure}[!h]
\centering
\includegraphics[width=0.41\linewidth,clip,trim=0 5 0 5]{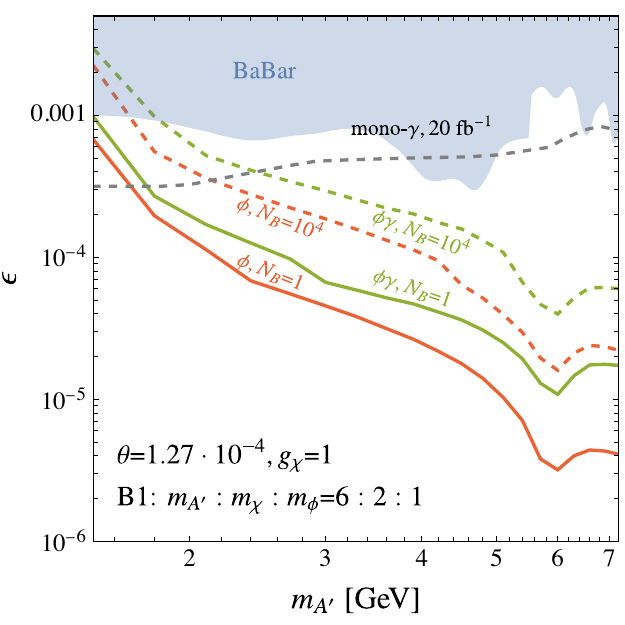}
\qquad
\includegraphics[width=0.41\linewidth,clip,trim=0 5 0 5]{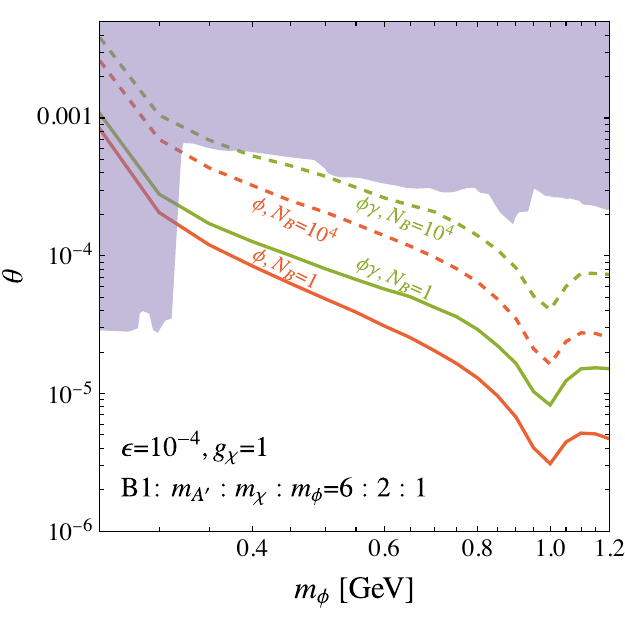}

\includegraphics[width=0.41\linewidth,clip,trim=0 5 0 5]{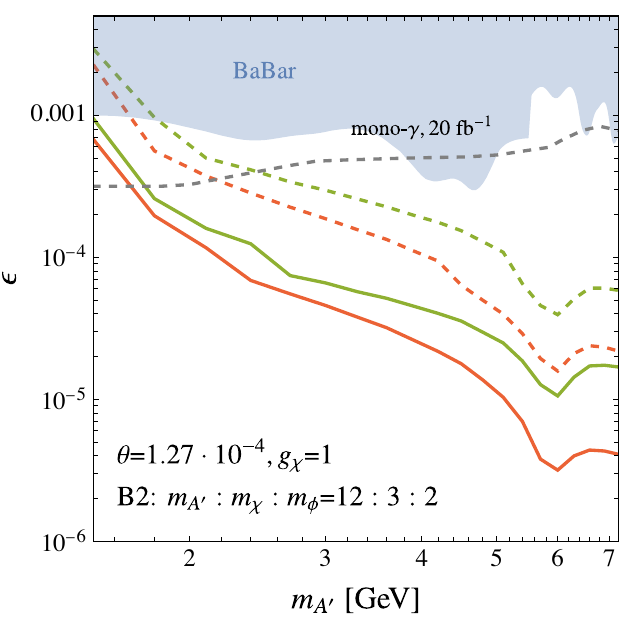}
\qquad
\includegraphics[width=0.41\linewidth,clip,trim=0 5 0 5]{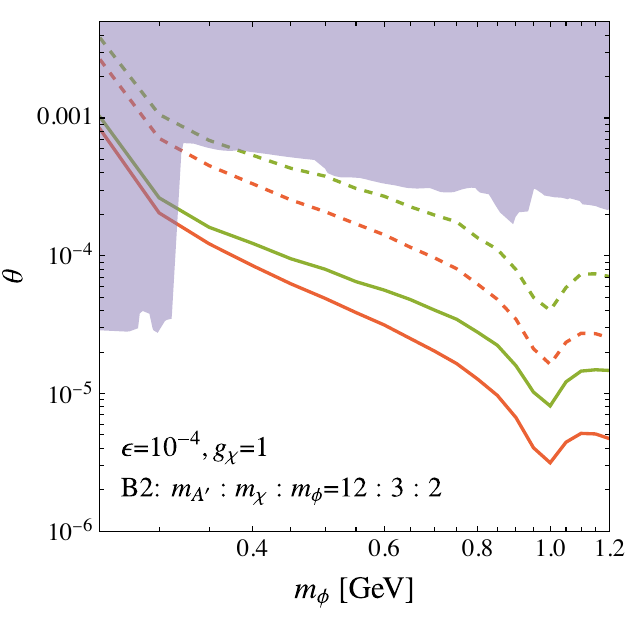}

\includegraphics[width=0.41\linewidth,clip,trim=0 5 0 5]{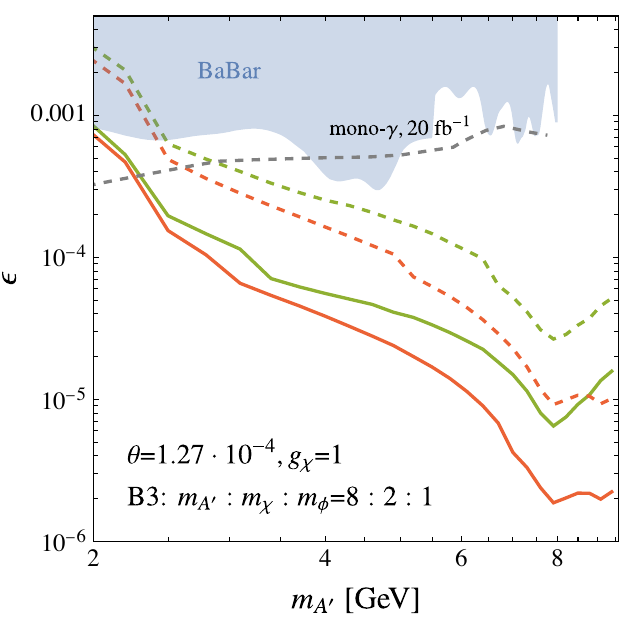}
\qquad
\includegraphics[width=0.41\linewidth,clip,trim=0 5 0 5]{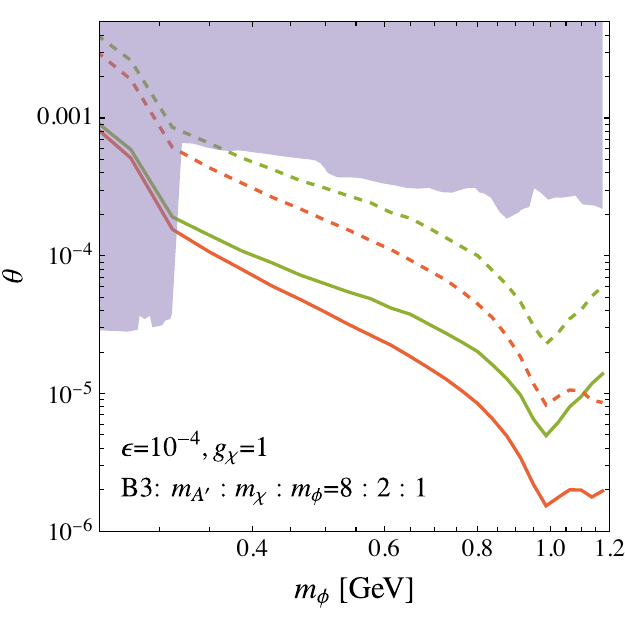}

\caption{Belle~II sensitivity to dark Higgs bosons $\phi$ with an integrated luminosity of ${\cal L}=50$ ab$^{-1}$ in terms of the kinetic mixing  $\epsilon$ (left column) or the Higgs mixing $\theta$ (right column). In each panel, the $\phi$ production channel is represented by red lines, the $\gamma \phi$ production channel by green lines. Solid lines corresponds to $N_b=1$ background event, dashed lines to $N_b=10^4$ events. In the left column, the blue shaded region is excluded by a BaBar search for $e^+ e^- \to \gamma A^\prime, A^\prime \to$ invisible~\cite{BaBar:2017tiz}; the grey dashed line is the corresponding Belle~II sensitivity projection for $\mathcal{L} = 20\,\mathrm{fb}^{-1}$~\cite{b2_phys_book}. In the right column, the purple area is a combination of exclusions taken from Ref.~\cite{Ferber:2023iso}.}
\label{fig:sensitivities}
\hspace{-5mm}
\end{figure}

Our results are presented in in \cref{fig:sensitivities}. We show the Belle~II sensitivity to both $\phi$ production (red lines) and $\gamma \phi$ production (green lines) with an integrated luminosity of ${\cal L}=50$ ab$^{-1}$ for three benchmark scenarios that keep the mass ratios fixed, see eqs.~\eqref{def:b1}--\eqref{def:b3}. The lower, solid lines (best sensitivity) corresponds to $N_b = 1$ background events in the $\phi$ channel, the upper, dashed lines (worst sensitivity) corresponds to $N_b = 10^4$ background events in the $\phi$ channel. In each case, the background level for the $\gamma \phi$ channel is obtained by a rescaling as discussed in section~\ref{sec:simulation}.
We fix $g_\chi = 1$ for concreteness, but emphasize that slightly larger or smaller values may be needed to reproduce the observed dark matter relic abundance as discussed in section~\ref{sec:model}.

In the left column, we have fixed $\theta = 1.27 \cdot 10^{-4}$ and vary the kinetic mixing parameter $\epsilon$, such that our results can be compared to existing bounds~\cite{BaBar:2017tiz} and projected sensitivities~\cite{b2_phys_book} from searches for invisibly decaying dark photons. This comparison is possible because in our model the dark photon can either decay invisibly ($A' \to \chi \bar{\chi}$) or semi-visibly ($A' \to \chi \bar{\chi} \phi$). Our results clearly demonstrate that including the latter decay mode leads to significantly improved sensitivity. 

In the right column we instead fix $\epsilon = 10^{-4}$ and present the sensitivities in terms of the Higgs mixing $\theta$, which is constrained by a multitude of experimental null results (see Ref.~\cite{Ferber:2023iso}). We emphasize, that these results rely exclusively on Higgs mixing for both production and decay of dark Higgs bosons, while in our set-up there exists an additional production mode via kinetically mixed dark photons. This additional production mode is the reason for the significant improvements in sensitivity that can be achieved.

The improvements in sensitivity are most substantial for large masses of the dark sector particles. To some degree, this is driven by the corresponding enhancement in the cross section as shown  in~\cref{fig:xsec_plot}. However, the main factor is the dark Higgs boson lifetime, which decreases with increasing dark Higgs boson mass. For the parameter regions under consideration, the dark Higgs boson decay length $c\tau_\phi$ is much larger than the maximal radius $r_\text{out}$ of the sensitive detector volume, such that only a small fraction of the produced dark Higgs bosons will decay within the detector. This is also the reason why the number of expected signal events (see the tables in appendix~\ref{app:efficiencies}) is much smaller than the naive expectation based on the total cross section given in \cref{fig:xsec_plot}.Using eqs.~\eqref{def:dec_prob} and~\eqref{def:mean_dist}, we can estimate this fraction as $(r_\text{out} - r_\text{in}) \Gamma_\phi m_\phi / p_i^T$. Assuming that $p_i^T$ depends mostly on the centre-of-mass energy and only mildly on the dark Higgs boson mass, it becomes clear that sensitivity increases for larger masses. Moreover, this estimate also shows that, while the dark Higgs production cross section is proportional to $\epsilon^2$ but independent of $\theta$, the predicted number of signal events is proportional to both $\epsilon^2$ and $\Gamma_\phi \propto \theta^2$. This finding explains the similarity of the sensitivity curves in both columns.

We note that in the right column, the sensitivity of the proposed search does not extend up to arbitrarily high values of $\theta$, because at some point the dark Higgs bosons would decay so quickly that the probability of a displaced vertex is exponentially suppressed. However, this effect is only relevant in the parameter region that is already excluded by other experiments. Within the allowed parameter region shown in \cref{fig:sensitivities}, the dark Higgs boson decay length is always larger than $r_\text{in}$. 

We furthermore observe that the $\phi$ production channel is always slightly more sensitive (by a factor of 2--3 in $\epsilon$ or $\theta$) than the $\gamma \phi$ production channel. This is a direct consequence of the larger production cross section for the former compared to the latter. However, our central result is that both channels may be observed simultaneously and can be correlated using the invariant mass of the particles produced in the decays of the dark Higgs boson as well as the distribution of $m_{\chi\chi}$ and $v$ as defined in eq.~\eqref{def:v_hyperbole}, which carry information about the dark photon mass.

\begin{figure}[t]
    \centering
    \includegraphics[width=\linewidth,clip,trim=40 15 40 15]{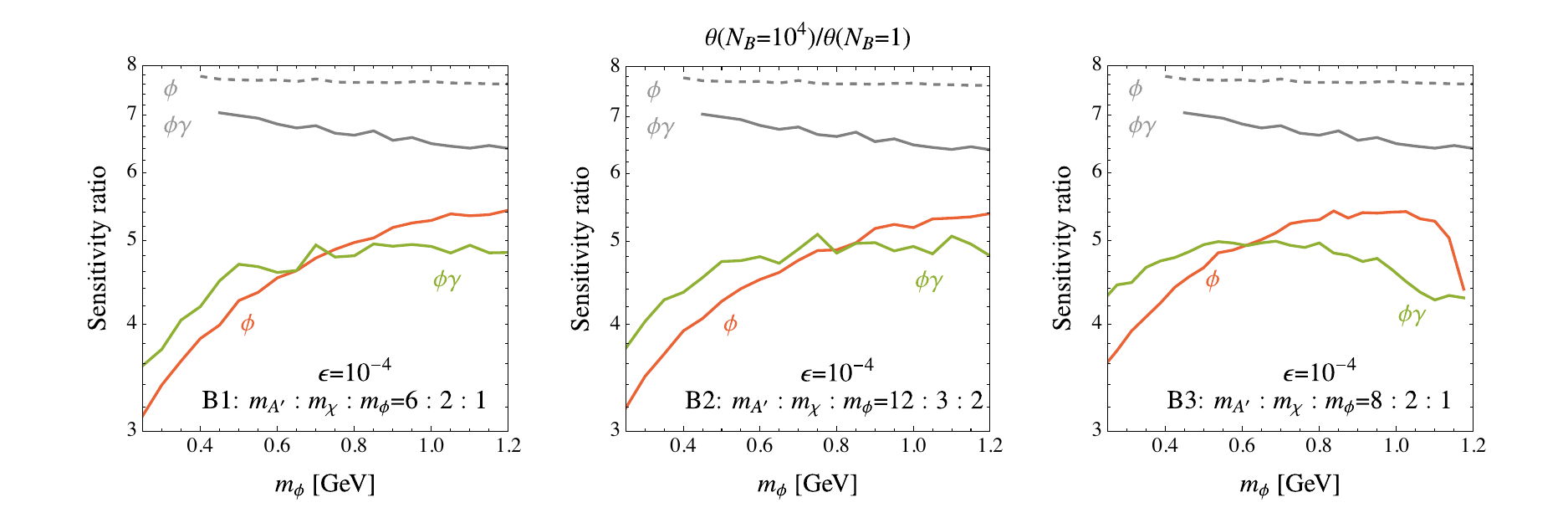}
    \caption{Ratio of the Belle~II sensitivity for dominant background ($N_B=10^4$) and negligble background ($N_B=1$). The red line corresponds to the $\phi$ production and matches the red band in \cref{fig:sensitivities}; the green line corresponds to the $\gamma \phi$ production and matches the green band in \cref{fig:sensitivities}; the gray lines indicate the sensitivity ratios expected for a simple cut-and-count analysis, i.e.\ without exploiting the differences in kinematic distributions of signal and background. }
    \label{fig:rat_reach}
\end{figure}

The characteristic distribution of signal events in $m_{\chi\chi}$ and $v$ are also the reason why the degradation in sensitivity for larger background is much milder than expected for a simple cut-and-count analysis. For a statistics limited search, one would expect that increasing backgrounds by a factor of $10^4$ should decrease the testable signal cross section by approximately a factor of $10^2$, corresponding to a loss in sensitivity by one order of magnitude in $\epsilon$ or $\theta$. Instead, we find that the loss in sensitivity for $N_b = 10^4$ is only a factor of around 3 (for small masses) to 5 (for high masses), see \cref{fig:rat_reach}. This improvement is a result of optimizing the selection cuts based on the number of background events, see \cref{fig:optim_algo_histos}.

Of course, for very large background rates, the actual sensitivity will depend on the precise shape of the background distribution (which we do not attempt to estimate accurately) as well as on possible systematic uncertainties.
Nevertheless, our results highlight the benefits of using suitable kinematic variables to improve the signal-to-background ratio.

Let us briefly comment on triggers. For our analysis, we have assumed that all signal events pass the trigger. This would certainly be the case for a dedicated displaced vertex trigger as proposed in Ref.~\cite{felix_trigger}. In the absence of such a trigger, we need to rely on the tracks from charged particles and energy depositions in the calorimeter. For the $\gamma \phi$ production process, there are two tracks coming from the dark Higgs boson decay as well as a single photon. From \cref{sec:distributions} we know that the signal prefers either configurations with large $E_\gamma$ or large $E_\phi$, such that we expect either a high energy deposition in the calorimeter or several energetic tracks, for both of which the existing triggers are expected to be very efficient. For the $\phi$ production process, on the other hand, the signal consists only of the two tracks coming from the dark Higgs boson decay. If $m_\phi$ is small and $m_{A'}$ (and hence $\mcc$ is large), the tracks may become so feeble that the events become more difficult to trigger. In such a case, our sensitivity estimates can only be fully realized once a displaced vertex trigger has been implemented.

To conclude, let us emphasize that all our results are based on the assumption that the ratio between the $\gamma \phi$ and the $\phi$ background rates is given by a constant factor 0.22. If the $\gamma \phi$ background turns out to be even smaller, this might invert the importance of the two channels for a discovery search. Moreover, if backgrounds for the $\gamma \phi$ search are found to be sufficiently small, an attractive possibility would be to extend the search to include also displaced vertices in the instrumented region between $17 \, \mathrm{cm}$ and $60 \, \mathrm{cm}$ away from the beam axis in the radial direction. While backgrounds in this region are likely non-negligible, the signal rate may increase sufficiently to allow for a further improvement in sensitivity.

\section{Conclusions}\label{sec:conclusions}

In this work we considered a dark sector consisting of a dark photon $A^\prime$, a fermionic dark matter particle $\chi$ and a dark Higgs boson $\phi$. The dark photon is the gauge boson of a dark $U(1)^\prime$ symmetry with gauge coupling $g_{\chi}$ and acts as the mediator between dark matter and SM particles, to which it couples via kinetic mixing of strength $\epsilon$. The mass of the dark photon is generated by a dark Higgs mechanism, and the corresponding dark Higgs boson couples to SM particles via the Higgs mixing parameter $\theta$. 

In our study we focused on the dark Higgs boson mass range $0.2 \, \mathrm{GeV} \leq m_\phi \leq 1.2 \, \mathrm{GeV}$ and considered the mass hierarchy $\mAp> 2 m_\chi + m_\phi$ and $m_\chi > m_\phi$, such that the decay $A^\prime \to \phi \chi \bar{\chi}$ is kinematically allowed, but the decay $\phi \to \chi \bar{\chi}$ is forbidden; specifically, we focus the mass benchmarks in \cref{def:b1,def:b2,def:b3}. The dark Higgs boson therefore always decays into SM particles, while the dark photon has an invisible ($A^\prime \to \chi \bar{\chi}$) and a semi-visible ($A^\prime \to \chi \bar{\chi} \phi$, $\phi \to$ SM) decay mode. Moreover, in the parameter regions of interest, the dark Higgs boson is typically long-lived, such that its decays give rise to a displaced vertex. 

In the early universe, dark matter particles can annihilate via the process $\chi \bar{\chi} \to \phi \phi$. Since in our set-up the dark matter particles do not directly couple to the dark Higgs boson, these annihilations arise only through dark photon loops. We compute the resulting cross section and dark matter relic abundance and find that for dark matter masses at the GeV scale, the measured value is reproduced for $g_{\chi}\sim 1$. For such large gauge couplings, the probability of dark Higgs-strahlung is sizeable, such that we can expect a large number of events with dark Higgs bosons at Belle~II.

In our study, we considered two new channels for the discovery of such a dark sectors at Belle~II, namely dark Higgs-strahlung with or without an additional photon from initial-state radiation, which we name $\gamma \phi$ and $\phi$ production, respectively. We focused specifically on the case that the dark Higgs boson decays into a pair of SM particles within the uninstrumented region of the detector between 2 and 9 mm from the interaction point in the radial direction. While this requirement is very restrictive and significantly reduces the expected signal, backgrounds from known SM processes are expected to be tiny or absent. 

But even if backgrounds turn out to be non-negligible in the very large data sets to be collected by Belle~II, the characteristic features of the signal will allow for a powerful rejection of background events. To demonstrate this approach, we considered a mock background simulation, in which we sampled uniformly from the phase spaces of the final-state particles, and left the total number of background events $N_B$ as a free parameter. While the background distribution is smooth, the signal exhibits peaks in certain kinematic variables resulting from kinematic configurations that give a resonant enhancement from the dark photon propagator(s). The $\phi$ production peaks around $\mcc \sim \mAp$, while the $\gamma \phi$ production peaks either around $\mcc \sim \mAp$ or around $E_\gamma \sim \bar{E}$. We combined both of these peaks into a new kinematic variable called $v$, defined in \cref{def:v_hyperbole}. These features are not expected in the background, allowing for an optimised signal selection as a function of the background level $N_B$. 

We then calculated the expected sensitivity for Belle~II in terms of $\epsilon$ and $\theta$ both in the case that background are negligible ($N_B = 1$) and that backgrounds are large ($N_B = 10^4$). We find that both $\phi$ and $\gamma \phi$ production predict observable signals in currently allowed parameter regions for all benchmarks that we consider, see \cref{fig:sensitivities}. While larger background rates reduce the sensitivity, the reduction is much smaller than expected for a naive cut-and-count analysis, thanks to the distinctive features of the signal. Hence, even in the case of non-negligible backgrounds, Belle~II offers tantalising prospects for the discovery of a dark sector featuring dark Higgs-strahlung.

\acknowledgments

We thank Michael Baker, Martin Bauer, Torben Ferber, Ana Foguel, Andrea Thamm and Jose Zurita for discussions and Patrick Ecker for valuable comments on the manuscript. FK acknowledges funding from the Deutsche Forschungsgemeinschaft
(DFG) through Grant No. 396021762 -- TRR 257. FA is
grateful to the Mainz Institute for Theoretical Physics (MITP) of the Cluster of Excellence PRISMA+ (Project ID 390831469), for
its hospitality and its partial support during the completion of this work.

\bibliographystyle{JHEP_improved}
\bibliography{bibliography}

\appendix

\section{Smearing and binning}\label{app:smear_bin}

\subsection{Smearing}\label{subsec:smearing}

For our analysis we do not run a detailed detector simulation. Instead, we model the resolution on the measurements on all final energies and masses with normal distributions centred at the truth-level value and standard deviations as given in Ref.~\cite{b2_phys_book}:
\begin{equation}
\frac{\delta E}{E}=\sqrt{
\left[\frac{0.066\%}{E/\GeV}\right]^2 + \left[ \frac{0.81\%}{(E/\GeV)^{1/4}}\right]^2 + \left[ 1.34\% \right]^2
}.\label{eq:smearing}
\end{equation}
In practice, every entry for $E_\gamma$ or $\mcc$ is substituted with an array of 100 values randomly extracted from a normal distribution. 

\subsection{Binning}\label{subsec:binning}

For $\gamma\phi$ production, the signal is very peaked at $v\sim0$, while the background is almost independent of $v$. For large $N_b$ it is therefore generally advantageous to restrict the search window to a single bin centred around 0. The sensible minimal width of this bin depends on the detector resolution and hence on the smearing discussed above. A good choice of binning is such that the number of events in a given bin before and after smearing are compatible with each other within Poisson fluctuations. In order to choose appropriate binning for $v$, it is therefore essential to study the effect of smearing on the distribution of $v$. Consider for example a measurement very close to the axes $E_\gamma \sim \bar{E}$ and/or $\mcc \sim \mAp$:
\begin{equation}
E_\gamma=\bar{E}-x \, , \quad \mcc=\mAp-y \; ,\label{eq:vxy}
\end{equation}
where we assume $x,y>0$, such that $v= x y > 0$. 
Say that after smearing 
\begin{equation}
E_\gamma=\bar{E}+x^\prime \, , \quad \mcc=\mAp-y^\prime\label{eq:vxy_prime}
\end{equation}
with  $x^\prime,y^\prime>0$, such that $v^\prime < 0$. The difference in $v$ is $\Delta v=|v-v^\prime|=xy+ x^\prime y^\prime$. The fact that $v$ can be both positive and negative increases the size of possible fluctuations, which implies a larger required bin size. Since the sign of $v$ does not carry relevant information, we can simply neglect it. Our variable of choice is therefore $\sqrt{|v|}$, so that \cref{eq:vxy,eq:vxy_prime} become
\begin{equation}
\sqrt{|v|}= \sqrt{x y}, \qquad \sqrt{|v^\prime|}= \sqrt{x^\prime y^\prime}, \qquad \Delta \sqrt{|v|}=|\sqrt{xy}- \sqrt{x^\prime y^\prime}| < \sqrt{\Delta v} \; .
\end{equation}

\section{Efficiency tables}
\label{app:efficiencies}

In this appendix we provide details on the event selection for Benchmark 1. For each value of the dark photon mass, we give the predicted number of signal events, the optimised selection cuts and their efficiencies for signal and background, as well as the resulting likelihoods.
Tables~\ref{tab:eff_no_a_1} and \ref{tab:eff_no_a_1e4} correspond to $\phi$ production for $N_b = 1$ and $N_b = 10^4$, respectively, while tables~\ref{tab:eff_a_1} and \ref{tab:eff_a_1e4} provide the corresponding information for $\gamma\phi$ production.

\begin{table}
\centering
\begin{tabular}{ccccccc}
\toprule
$\mAp$ [GeV]& Resc. Sig & $ \mcc^\text{min}$ [GeV]& $\mcc^\text{max}$ [GeV]& Select. Sig & Select. Bg & Log-Likelihood\\ 
\midrule
$1.2$ & $4.6\times 10^{-6}$ & $1.2$ & $1.2$ & $3.3\times 10^{-6}$ & $3.1\times 10^{-3}$ & $3.5\times 10^{-9}$ \\ $
 1.5$ & $5.1\times 10^{-2}$ & $1.5$ & $1.6$ & $4.3\times 10^{-2}$ & $7.1\times 10^{-3}$ & $5.8\times 10^{-2}$ \\ $
 1.8$ & $8.5\times 10^{-1}$ & $1.6$ & $2.$ & $8.\times 10^{-1}$ & $2.2\times 10^{-2}$ & $1.4$ \\ $
 2.1$ & $2.5$ & $1.8$ & $2.4$ & $2.4$ & $4.1\times 10^{-2}$ & $4.6$ \\ $
 2.4$ & $5.2$ & $2.$ & $2.8$ & $5.1$ & $6.7\times 10^{-2}$ & $9.7$ \\ $
 2.7$ & $9.6$ & $2.2$ & $3.2$ & $9.4$ & $1.\times 10^{-1}$ & $1.8\times 10^1$ \\ $
 3.$ & $1.6\times 10^1$ & $2.3$ & $3.7$ & $1.6\times 10^1$ & $1.5\times 10^{-1}$ & $3.1\times 10^1$ \\ $
 3.3$ & $2.6\times 10^1$ & $2.3$ & $4.3$ & $2.6\times 10^1$ & $2.3\times 10^{-1}$ & $5.\times 10^1$ \\ $
 3.6$ & $4.2\times 10^1$ & $2.4$ & $4.8$ & $4.2\times 10^1$ & $2.9\times 10^{-1}$ & $8.\times 10^1$ \\ $
 3.9$ & $6.3\times 10^1$ & $2.6$ & $5.2$ & $6.2\times 10^1$ & $3.5\times 10^{-1}$ & $1.2\times 10^2$ \\ $
 4.2$ & $9.7\times 10^1$ & $2.7$ & $5.8$ & $9.7\times 10^1$ & $4.3\times 10^{-1}$ & $1.9\times 10^2$ \\ $
 4.5$ & $1.5\times 10^2$ & $2.9$ & $6.2$ & $1.5\times 10^2$ & $4.8\times 10^{-1}$ & $3.\times 10^2$ \\ $
 4.8$ & $2.5\times 10^2$ & $3.2$ & $6.5$ & $2.5\times 10^2$ & $5.2\times 10^{-1}$ & $4.9\times 10^2$ \\ $
 5.1$ & $4.4\times 10^2$ & $3.4$ & $7.$ & $4.4\times 10^2$ & $5.9\times 10^{-1}$ & $8.8\times 10^2$ \\ $
 5.4$ & $9.\times 10^2$ & $3.5$ & $7.5$ & $9.\times 10^2$ & $6.5\times 10^{-1}$ & $1.8\times 10^3$ \\ $
 5.7$ & $2.3\times 10^3$ & $3.8$ & $8.$ & $2.3\times 10^3$ & $7.1\times 10^{-1}$ & $4.6\times 10^3$ \\ $
 6.$ & $3.7\times 10^3$ & $3.9$ & $8.4$ & $3.7\times 10^3$ & $7.6\times 10^{-1}$ & $7.4\times 10^3$ \\ $
 6.3$ & $2.\times 10^3$ & $4.2$ & $8.8$ & $2.\times 10^3$ & $7.9\times 10^{-1}$ & $4.\times 10^3$ \\ $
 6.6$ & $1.6\times 10^3$ & $4.3$ & $9.2$ & $1.6\times 10^3$ & $8.2\times 10^{-1}$ & $3.1\times 10^3$ \\ $
 6.9$ & $1.7\times 10^3$ & $4.5$ & $9.6$ & $1.7\times 10^3$ & $8.3\times 10^{-1}$ & $3.3\times 10^3$ \\ $
 7.2$ & $1.9\times 10^3$ & $4.8$ & $9.6$ & $1.9\times 10^3$ & $7.9\times 10^{-1}$ & $3.8\times 10^3$ \\ $
 7.5$ & $2.5\times 10^3$ & $5.2$ & $9.6$ & $2.5\times 10^3$ & $7.4\times 10^{-1}$ & $5.\times 10^3$ \\
 \bottomrule
\end{tabular}
\caption{Efficiency table for events from Benchmark 1 for the process $e^- e^+ \to \phi \chi \bar{\chi}$. The simulation parameters are: $\epsilon=10^{-4}$, $\theta=10^{-4}$, $N_b=1$. The mass points are identified by the $A^\prime$ mass in the first column. All events numbers assume 50 ab$^{-1}$ integrated luminosity. The second column gives the number of signal events after applying the displacement probability in \cref{def:dec_prob} but before selections.  The third and fourth column summarize the selection: $\mcc^\text{min} <\mcc< \mcc^\text{max}$.  The fifth column is the number of signal events after selection. The sixth column is the number of background events after selection. The last column is the log-likelihood computed on the selected signal and background.}
\label{tab:eff_no_a_1}
\end{table}
\begin{table}
\centering
\begin{tabular}{ccccccc}
\toprule
$\mAp$ [GeV]& Resc. Sig & $\mcc^\text{min}$ [GeV]& $\mcc^\text{max}$ [GeV]& Select. Sig & Select. Bg & Log-Likelihood\\ 
\midrule
$1.2$ & $4.6\times 10^{-6}$ & $1.2$ & $1.2$ & $3.3\times 10^{-6}$ & $3.1\times 10^1$ & $3.5\times 10^{-13} $ \\ $
 1.5$ & $5.1\times 10^{-2}$ & $1.5$ & $1.5$ & $3.7\times 10^{-2}$ & $4.8\times 10^1$ & $2.9\times 10^{-5} $ \\ $
 1.8$ & $8.5\times 10^{-1}$ & $1.8$ & $1.9$ & $6.7\times 10^{-1}$ & $8.3\times 10^1$ & $5.3\times 10^{-3} $ \\ $
 2.1$ & $2.5$ & $2.1$ & $2.2$ & $1.8$ & $9.1\times 10^1$ & $3.6\times 10^{-2} $ \\ $
 2.4$ & $5.2$ & $2.4$ & $2.5$ & $3.8$ & $1.2\times 10^2$ & $1.1\times 10^{-1} $ \\ $
 2.7$ & $9.6$ & $2.6$ & $2.8$ & $6.9$ & $1.6\times 10^2$ & $2.9\times 10^{-1} $ \\ $
 3.$ & $1.6\times 10^1$ & $2.9$ & $3.1$ & $1.2\times 10^1$ & $2.1\times 10^2$ & $6.2\times 10^{-1} $ \\ $
 3.3$ & $2.6\times 10^1$ & $3.2$ & $3.4$ & $1.8\times 10^1$ & $2.3\times 10^2$ & $1.3 $ \\ $
 3.6$ & $4.2\times 10^1$ & $3.5$ & $3.7$ & $2.9\times 10^1$ & $2.9\times 10^2$ & $2.8 $ \\ $
 3.9$ & $6.3\times 10^1$ & $3.8$ & $4.$ & $4.5\times 10^1$ & $3.4\times 10^2$ & $5.5 $ \\ $
 4.2$ & $9.7\times 10^1$ & $4.1$ & $4.3$ & $7.\times 10^1$ & $3.9\times 10^2$ & $1.1\times 10^1 $ \\ $
 4.5$ & $1.5\times 10^2$ & $4.4$ & $4.6$ & $1.2\times 10^2$ & $5.5\times 10^2$ & $2.2\times 10^1 $ \\ $
 4.8$ & $2.5\times 10^2$ & $4.7$ & $4.9$ & $1.7\times 10^2$ & $4.8\times 10^2$ & $5.1\times 10^1 $ \\ $
 5.1$ & $4.4\times 10^2$ & $5.$ & $5.2$ & $3.2\times 10^2$ & $5.2\times 10^2$ & $1.4\times 10^2 $ \\ $
 5.4$ & $9.\times 10^2$ & $5.2$ & $5.6$ & $7.4\times 10^2$ & $8.8\times 10^2$ & $4.\times 10^2 $ \\ $
 5.7$ & $2.3\times 10^3$ & $5.5$ & $5.9$ & $1.8\times 10^3$ & $7.9\times 10^2$ & $1.7\times 10^3 $ \\ $
 6.$ & $3.7\times 10^3$ & $5.8$ & $6.2$ & $3.1\times 10^3$ & $1.\times 10^3$ & $3.3\times 10^3 $ \\ $
 6.3$ & $2.\times 10^3$ & $6.1$ & $6.6$ & $1.7\times 10^3$ & $1.1\times 10^3$ & $1.3\times 10^3 $ \\ $
 6.6$ & $1.6\times 10^3$ & $6.4$ & $6.9$ & $1.3\times 10^3$ & $1.1\times 10^3$ & $8.8\times 10^2 $ \\ $
 6.9$ & $1.7\times 10^3$ & $6.7$ & $7.2$ & $1.4\times 10^3$ & $1.2\times 10^3$ & $9.3\times 10^2 $ \\ $
 7.2$ & $1.9\times 10^3$ & $7.$ & $7.4$ & $1.5\times 10^3$ & $9.2\times 10^2$ & $1.2\times 10^3 $ \\ $
 7.5$ & $2.5\times 10^3$ & $7.2$ & $7.8$ & $2.1\times 10^3$ & $1.2\times 10^3$ & $1.7\times 10^3 $ \\
 \bottomrule
\end{tabular}
\caption{Efficiency table for events from Benchmark 1 for the process $e^- e^+ \to \phi \chi \bar{\chi}$. The simulation parameters are: $\epsilon=10^{-4}$, $\theta=10^{-4}$, $N_b=10^4$. The mass points are identified by the $A^\prime$ mass in the first column. All events numbers assume 50 ab$^{-1}$ integrated luminosity. The second column gives the number of signal events after applying the displacement probability in \cref{def:dec_prob} but before selections.  The third and fourth column summarize the selection: $\mcc^\text{min} <\mcc< \mcc^\text{max}$.  The fifth column is the number of signal events after selection. The sixth column is the number of background events after selection. The last column is the log-likelihood computed on the selected signal and background.}
\label{tab:eff_no_a_1e4}
\end{table}
\begin{table}[!h]
\centering
\begin{tabular}{cccccc}
\toprule
$\mAp$ [GeV] & Init. Sig & $|v^\text{max}|$ [GeV$^2$]& Select. Sig & Select Bg & Log-Likelihood\\
\midrule
$1.2$ & $3.4\times 10^{-6}$ & $0.21$ & $2.3\times 10^{-6}$ & $4.3\times 10^{-3}$ & $1.3\times 10^{-9}$ \\
$1.5$ & $3.3\times 10^{-2}$ & $0.21$ & $2.1\times 10^{-2}$ & $5.2\times 10^{-3}$ & $2.6\times 10^{-2}$ \\
$1.8$ & $0.49$ & $0.35$ & $0.4$ & $1.3\times 10^{-2}$ & $0.7$ \\
$2.1$ & $1.3$ & $0.77$ & $1.2$ & $4.6\times 10^{-2}$ & $2.2$ \\
$2.4$ & $2.3$ & $0.77$ & $2.3$ & $5.\times 10^{-2}$ & $4.1$ \\
$2.7$ & $3.9$ & $0.77$ & $3.6$ & $5.4\times 10^{-2}$ & $6.8$ \\
$3.$ & $5.8$ & $0.99$ & $5.7$ & $8.\times 10^{-2}$ & $11.$ \\
$3.3$ & $8.3$ & $0.99$ & $8.2$ & $8.4\times 10^{-2}$ & $16.$ \\
$3.6$ & $12.$ & $0.99$ & $11.$ & $8.8\times 10^{-2}$ & $22.$ \\
$3.9$ & $16.$ & $0.99$ & $15.$ & $9.1\times 10^{-2}$ & $30.$ \\
$4.2$ & $22.$ & $1.3$ & $22.$ & $0.13$ & $42.$ \\
$4.5$ & $30.$ & $1.3$ & $30.$ & $0.13$ & $58.$ \\
$4.8$ & $45.$ & $1.3$ & $44.$ & $0.13$ & $86.$ \\
$5.1$ & $72.$ & $1.7$ & $71.$ & $0.18$ & $1.4\times 10^2$ \\
$5.4$ & $1.2\times 10^2$ & $3.7$ & $1.2\times 10^2$ & $0.22$ & $2.4\times 10^2$ \\
$5.7$ & $2.7\times 10^2$ & $2.8$ & $2.7\times 10^2$ & $0.22$ & $5.3\times 10^2$ \\
$6.$ & $3.7\times 10^2$ & $2.8$ & $3.7\times 10^2$ & $0.22$ & $7.5\times 10^2$ \\
$6.3$ & $2.1\times 10^2$ & $2.8$ & $2.1\times 10^2$ & $0.22$ & $4.2\times 10^2$ \\
$6.6$ & $1.5\times 10^2$ & $2.2$ & $1.5\times 10^2$ & $0.21$ & $3.\times 10^2$ \\
$6.9$ & $1.5\times 10^2$ & $1.7$ & $1.5\times 10^2$ & $0.18$ & $2.9\times 10^2$ \\
$7.2$ & $1.5\times 10^2$ & $1.7$ & $1.5\times 10^2$ & $0.18$ & $3.\times 10^2$ \\
$7.5$ & $1.6\times 10^2$ & $1.7$ & $1.6\times 10^2$ & $0.17$ & $3.3\times 10^2$ \\
\bottomrule
\end{tabular}
\caption{
Efficiency table for events from Benchmark 1 for the process $e^- e^+ \to \gamma \phi \chi \bar{\chi}$. The simulation parameters are: $\epsilon=10^{-4}$, $\theta=10^{-4}$, $N_b=1$. The mass points are identified by the $A^\prime$ mass in the first column. All events numbers assume 50 ab$^{-1}$ integrated luminosity. The second column gives the number of signal events after applying the displacement probability in \cref{def:dec_prob} but before selections.  The third column summarizes the selection: $|v|< |v^\text{max}|$. The fourth column is the number of signal events after selection. The fifth column is the number of background events after selection. The last column is the log-likelihood computed on the selected signal and background.}
\label{tab:eff_a_1}
\end{table}
\begin{table}[!h]
\centering
\begin{tabular}{cccccc}
\toprule
$\mAp$ [GeV] & Init. Sig & $|v^\text{max}|$ [GeV$^2$]& Select. Sig & Select Bg & Log-Likelihood  \\ 
\midrule
$1.2$ & $3.4\times 10^{-6}$ & $0.21$ & $2.3\times 10^{-6}$ & $43.$ & $1.3\times 10^{-13}$ \\
$1.5$ & $3.3\times 10^{-2}$ & $0.21$ & $2.1\times 10^{-2}$ & $52.$ & $8.7\times 10^{-6}$ \\
$1.8$ & $0.49$ & $0.21$ & $0.3$ & $58.$ & $1.5\times 10^{-3}$ \\
$2.1$ & $1.3$ & $0.27$ & $0.85$ & $99.$ & $7.2\times 10^{-3}$ \\
$2.4$ & $2.3$ & $0.27$ & $1.5$ & $1.1\times 10^2$ & $2.\times 10^{-2}$ \\
$2.7$ & $3.9$ & $0.27$ & $2.4$ & $1.2\times 10^2$ & $4.8\times 10^{-2}$ \\
$3.$ & $5.8$ & $0.27$ & $3.3$ & $1.2\times 10^2$ & $9.1\times 10^{-2}$ \\
$3.3$ & $8.3$ & $0.27$ & $4.5$ & $1.2\times 10^2$ & $0.16$ \\
$3.6$ & $12.$ & $0.27$ & $6.2$ & $1.3\times 10^2$ & $0.29$ \\
$3.9$ & $16.$ & $0.35$ & $10.$ & $1.9\times 10^2$ & $0.51$ \\
$4.2$ & $22.$ & $0.35$ & $13.$ & $1.9\times 10^2$ & $0.88$ \\
$4.5$ & $30.$ & $0.35$ & $17.$ & $1.9\times 10^2$ & $1.5$ \\
$4.8$ & $45.$ & $0.35$ & $24.$ & $1.8\times 10^2$ & $3.$ \\
$5.1$ & $72.$ & $0.45$ & $46.$ & $2.8\times 10^2$ & $6.9$ \\
$5.4$ & $1.2\times 10^2$ & $0.45$ & $79.$ & $2.7\times 10^2$ & $19.$ \\
$5.7$ & $2.7\times 10^2$ & $0.45$ & $1.6\times 10^2$ & $2.7\times 10^2$ & $72.$ \\
$6.$ & $3.7\times 10^2$ & $0.59$ & $2.8\times 10^2$ & $4.1\times 10^2$ & $1.3\times 10^2$ \\
$6.3$ & $2.1\times 10^2$ & $0.45$ & $1.2\times 10^2$ & $2.5\times 10^2$ & $48.$ \\
$6.6$ & $1.5\times 10^2$ & $0.59$ & $1.1\times 10^2$ & $3.8\times 10^2$ & $27.$ \\
$6.9$ & $1.5\times 10^2$ & $0.59$ & $1.1\times 10^2$ & $3.7\times 10^2$ & $26.$ \\
$7.2$ & $1.5\times 10^2$ & $0.59$ & $1.1\times 10^2$ & $3.5\times 10^2$ & $28.$ \\
$7.5$ & $1.6\times 10^2$ & $0.59$ & $1.2\times 10^2$ & $3.4\times 10^2$ & $34.$ \\
\bottomrule
\end{tabular}
\caption{Efficiency table for events from Benchmark 1 for the process $e^- e^+ \to \gamma \phi \chi \bar{\chi}$. The simulation parameters are: $\epsilon=10^{-4}$, $\theta=10^{-4}$, $N_b=10^4$. The mass points are identified by the $A^\prime$ mass in the first column. All events numbers assume 50 ab$^{-1}$ integrated luminosity. The second column gives the number of signal events after applying the displacement probability in \cref{def:dec_prob} but before selections.  The third column summarizes the selection: $|v|< |v^\text{max}|$. The fourth column is the number of signal events after selection. The fifth column is the number of background events after selection. The last column is the log-likelihood computed on the selected signal and background.}
\label{tab:eff_a_1e4}
\end{table}

\end{document}